\documentclass[journal=cmatex,manuscript=article]{achemso}

\usepackage[version=3]{mhchem}
\usepackage{url}
\usepackage[symbol]{footmisc}
\usepackage{float}
\usepackage{setspace}

\usepackage{lineno}

\newcommand*\zeo[1]{\textbf{#1}}

\author{Daniel Schwalbe-Koda}
\affiliation{Lawrence Livermore National Laboratory, Livermore, CA, United States}
\email{dskoda@llnl.gov}

\author{Daniel E. Widdowson}
\affiliation{University of Liverpool, Liverpool, United Kingdom}

\author{Tuan Anh Pham}
\affiliation{Lawrence Livermore National Laboratory, Livermore, CA, United States}

\author{Vitaliy A. Kurlin}
\affiliation{University of Liverpool, Liverpool, United Kingdom}
\email{vitaliy.kurlin@gmail.com}

\title{Inorganic synthesis-structure maps in zeolites with machine learning and crystallographic distances}

\begin{document}


\begin{abstract}
Zeolites are inorganic materials known for their diversity of applications, synthesis conditions, and resulting polymorphs.
Although their synthesis is controlled both by inorganic and organic synthesis conditions, computational studies of zeolite synthesis have focused mostly on organic template design.
In this work, we use a strong distance metric between crystal structures and machine learning (ML) to create inorganic synthesis maps in zeolites.
Starting with 253 known zeolites, we show how the continuous distances between frameworks reproduce inorganic synthesis conditions from the literature without using labels such as building units.
An unsupervised learning analysis shows that neighboring zeolites according to our metric often share similar inorganic synthesis conditions, even in template-based routes.
In combination with ML classifiers, we find synthesis-structure relationships for 14 common inorganic conditions in zeolites, namely Al, B, Be, Ca, Co, F, Ga, Ge, K, Mg, Na, P, Si, and Zn.
By explaining the model predictions, we demonstrate how (dis)similarities towards known structures can be used as features for the synthesis space.
Finally, we show how these methods can be used to predict inorganic synthesis conditions for unrealized frameworks in hypothetical databases and interpret the outcomes by extracting local structural patterns from zeolites.
In combination with template design, this work can accelerate the exploration of the space of synthesis conditions for zeolites.
\end{abstract}

\section{Introduction}

Zeolites are inorganic porous materials widely recognized for their rich polymorphism and numerous applications.\cite{Davis2002OrderedPorous,Cejka2007IntroductionZeolite,Vermeiren2009ImpactZeolites}
Their porous structure provides unique opportunities to tailor materials performance in catalysis, gas adsorption, selective membranes, and more.\cite{Li2017ApplicationsZeolites,Li2021EmergingApplications,Dusselier2018SmallporeZeolites}
In principle, the performance of zeolites for each application can be controlled by adequate selection of polymorph and composition.
However, this selection is often hindered by the high-dimensional synthesis routes required to produce the materials \cite{Corma2004RationalizationZeolite}.
Zeolites are often synthesized with hydrothermal treatments, with inorganic and organic precursors cooperating to crystallize the nanoporous structure \cite{Cundy2003HydrothermalSynthesis}.
Organic templates are known to direct the formation of certain topologies, thus biasing the phase competition landscape to favor the best-matching topology instead of another \cite{Lobo1995StructuredirectionZeolite,Cundy2003HydrothermalSynthesis}.
Because of this effect, design of organic structure-directing agents (OSDAs) led to multiple successful examples of phase-selective zeolite synthesis and control of catalytic properties \cite{Barrer1981ZeolitesTheir,Brand2017EnantiomericallyEnriched,Gallego2017InitioSynthesis}, especially when used in combination with computational methods \cite{Lewis1996DeNovo,Sastre2005SearchingOrganic,Schmidt2015ComputationallyguidedSynthesis,Davis2016ComputationallyGuided,Schwalbe-Koda2021PrioriControl,schwalbe2021data,Bello2022TunableCHA}.

On the other hand, computational design of inorganic synthesis conditions for zeolites has not yet achieved the same impact as template design.
Despite their promise in controlling active site distribution \cite{DiIorio2020CooperativeCompetitive}, phase selectivity \cite{Shin2019RediscoveryImportance}, Si/Al ratio \cite{Li2023MachineLearningassisted}, morphology \cite{Li2019StrategiesControl}, or lowering the cost of syntheses \cite{Lee2021SynthesisThermally}, selection of inorganic conditions capable of synthesizing existing and novel zeolites is not easily modeled \cite{Oleksiak2014SynthesisZeolites}.
Recent progress in quantifying the role of inorganic synthesis conditions in zeolites includes: coupling machine learning and literature extraction \cite{Jensen2019MachineLearning,Muraoka2019LinkingSynthesis}; obtaining structure-synthesis correlations from synthesis routes \cite{Shin2019RediscoveryImportance,Asselman2022StructuralAspects}; predicting effects of inorganic cations in heteroatom distributions \cite{DiIorio2020CooperativeCompetitive,Schwalbe-Koda2021PrioriControl}; or using ML to control composition and particle sizes from template-free syntheses \cite{Li2023MachineLearningassisted}.
Nevertheless, their reliance on reported data prevents them to propose inorganic conditions for the synthesis of novel or hypothetical frameworks.
Whereas some inorganic synthesis-structure relationships can be derived from building units \cite{Itabashi2012WorkingHypothesis,Li2015SynthesisNew,Muraoka2019LinkingSynthesis} or alternative structural descriptors \cite{Asselman2022StructuralAspects}, automatically screening for new structures in hypothetical zeolite databases requires bypassing human-crafted labels such as building units.
Furthermore, although graph-theoretical methods can detect composite building units (CBUs) in arbitrary structures, their computational cost may be prohibitive when exploring large datasets.
Data-driven methods based in the topology of the structure also provide information on key factors that govern  kinetics of zeolite crystallization \cite{Blatov2013ZeoliteConundrum,Kuznetsova2018PredictingNew}, but do not immediately inform their synthesis conditions.
Finally, aggregate framework information such as density-energy plots \cite{Pophale2011DatabaseNew,Li2019NecessityHeteroatoms} or local interatomic distances \cite{Li2013CriteriaZeolite} provide few correlations between different inorganic synthesis conditions and targeted frameworks, which motivate new data-driven approaches to synthesizability prediction \cite{Helfrecht2019NewKind}.
Thus, advancing towards \textit{a priori} discovery of novel zeolite frameworks requires developing methods to: (1) uncover synthesis-structure relationships in zeolites; (2) efficiently explore the inorganic synthesis space of zeolites; and (3) bypass the absence of labeled data in hypothetical zeolite databases.

\begin{figure}[!h]
    \centering
    \includegraphics[width=\linewidth]{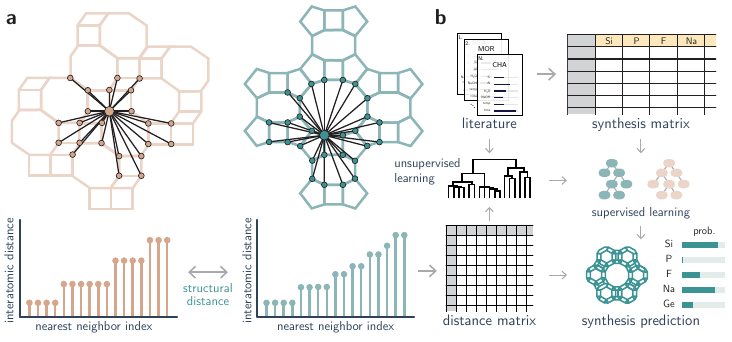}
    \caption{\textbf{Computational methods used to extract relationships between zeolite structures and their associated inorganic synthesis conditions.} \textbf{a}, Using the concept of AMD and the distance between these invariants, we compute a distance matrix between known zeolites. \textbf{b}, This information is combined with literature data and ML methods to correlate structural patterns with inorganic conditions.}
    \label{fig:01-methods}
\end{figure}

In this work, we correlate inorganic synthesis conditions to zeolite structure using a mathematically strong representation for comparing periodic crystals, the Average Minimum Distance (AMD)\cite{widdowson2022average}, derived from the Pointwise Distance Distribution (PDD)\cite{widdowson2022resolving} of crystal structures (see Fig. \ref{fig:01-methods}).
The PDD is independent of a unit cell, continuous under small perturbations, theoretically complete for generic crystals, and distinguished all periodic crystals in the Cambridge Structural Database.
Importantly, it can be computed much more efficiently than graph-based approaches, only requiring a fast nearest neighbor search \cite{elkin2023new}.
We show that the AMD vectors of zeolites can be used to predict inorganic synthesis conditions and recall a comprehensive dataset of synthesis conditions from the literature.
Then, we demonstrate that unsupervised and supervised machine learning (ML) methods can be used to create structure-synthesis relationships independently from OSDA design.
Finally, we propose inorganic synthesis conditions to realize hypothetical frameworks based on distances toward structures whose synthesis is known, thus proposing interpretable synthesis-structure models to guide the synthesis of new zeolites.

\section{Results and Discussion}

\subsection{Inorganic synthesis maps from unsupervised learning}

Designing synthesis-structure relationships in zeolites has long relied on intuitive patterns emerging from their natural structure.
For example, some CBUs are typically known to be synthesized by different inorganic conditions, such as \textit{d6r} in the presence of sodium ions or \textit{d4r} in the presence of germanium or fluorine.
Nevertheless, not all structures produced with certain inorganic agents exhibit the same CBUs, and CBUs are not necessarily realized only by one element.
Data-driven methods showed promise in connecting zeolite synthesizability to their local structure \cite{Li2013CriteriaZeolite} or accelerating their screening \cite{Helfrecht2019NewKind}, but interpreting large databases of structures can be challenging depending on the selected data representations.
For instance, subgraph isomorphism used to extract building units is an NP-type problem, and requires creating a predefined set of substructures that are being searched for.
On the other hand, local structural fingerprints such as the Smooth Overlap of Atomic Positions (SOAP) \cite{Bartok2013OnRepresenting} may be unable to distinguish between certain atomic environments \cite{pozdnyakov2020incompleteness}.

\begin{figure}[!h]
    \centering
    \includegraphics[width=\linewidth]{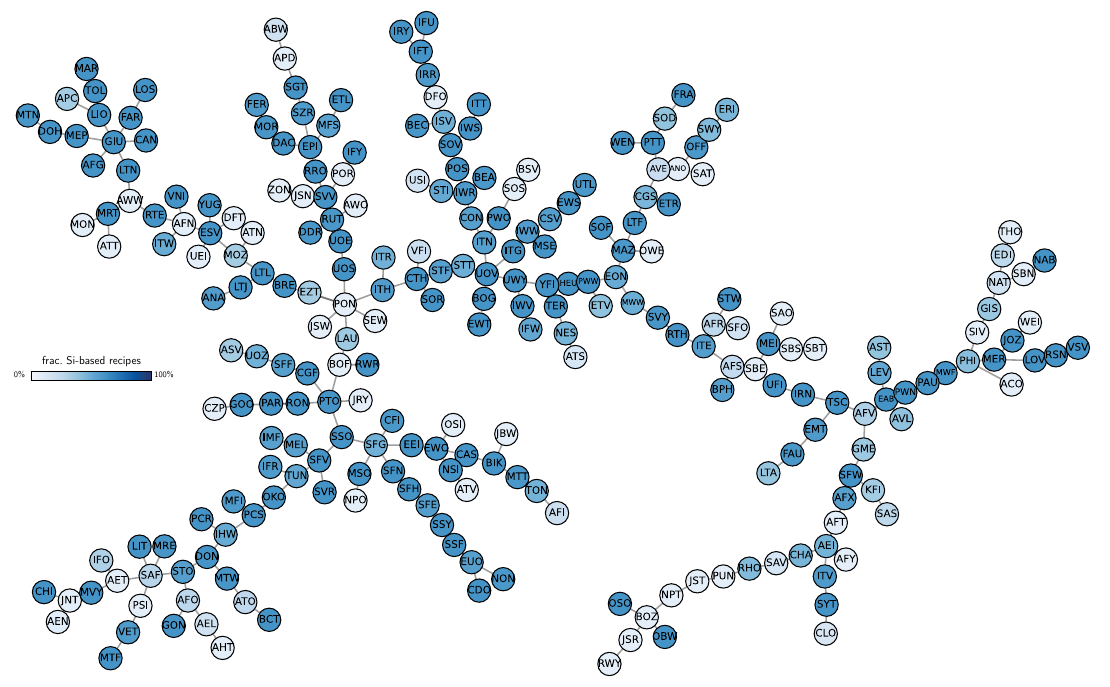}
    \caption{\textbf{Minimum spanning tree of 253 zeolites in the IZA database according to the $L_\infty$ metric on AMD vectors of length $k=100$ atomic neighbors}. Each framework is a node, and edges minimize the total length of a tree. Darker (lighter) colors indicate that silicon is more (less) frequent in the synthesis of each zeolite.}
    \label{fig:02-mst}
\end{figure}

To address these problems, we created a procedure to compare zeolites and extract synthesis-structure relationships without relying on any structural labels except for the atomic positions (Fig. \ref{fig:01-methods}).
Using the concept of PDDs and AMDs, we calculated the distance between two zeolite structures by comparing their AMD (see Methods).
This comparison between average local environments is computationally efficient but descriptive enough to distinguish all crystals in the CSD \cite{widdowson2022resolving}.
Then, we assumed that zeolites sharing similar local structures exhibit similar inorganic synthesis conditions.
To test this hypothesis, we computed the distance matrix between 253 known frameworks in the International Zeolite Association (IZA) database using AMD, and then performed a qualitative analysis of the results.
We found that the AMDs correlated weakly with differences of density and with the SOAP distance between structures, but showed almost no correlation with graph-based distances from previous work \cite{Schwalbe-Koda2019GraphSimilarity} (Fig. \ref{fig:si:01-analysis-distances}).
Moreover, we noted that zeolites sharing the lowest distances according to our metric have often been synthesized together (see Table \ref{tab:si:iza_pairs} in the Supporting Information).
Recovering pairs of structurally similar frameworks such as \zeo{ITH}-\zeo{ITR}, \zeo{SBS}-\zeo{SBT}, or \zeo{MWF}-\zeo{PAU} at low distance already suggests that the similarity metric is qualitatively sound.
To generalize this observation, we charted a map of zeolite structures based on their distances.
Figure \ref{fig:02-mst} shows the minimum spanning tree created by converting the AMD distance matrix into a graph with weighted edges.
Although the tree shows discrete connections and may not be accurate in the presence of outliers, it facilitates a qualitative interpretation of the results and may provide insights about synthesis-structure maps.
Even without considering synthesis labels of the data in Fig. \ref{fig:02-mst}, known relationships between zeolites emerge naturally from the structural tree map.
Zeolites known for their similar building patterns are clustered together in the minimum spanning tree, demonstrating that their AMD values capture the space of zeolites without learnable features.
Examples of such clusters include the ABC-6 zeolites, structures containing \textit{lov} building units, six-membered rings frameworks (e.g., \zeo{GIU} cluster), Ge- or boron-containing zeolites (e.g., \zeo{BEC} or \zeo{IRR} and \zeo{SFN}-\zeo{SSF} branch, respectively), to name a few (see also Fig. \ref{fig:si:02-mst-guide} for a visual guide).
Similarly, structural outliers such as the low-density \zeo{RWY} or \zeo{JSR}, or interrupted frameworks such as \zeo{-CLO}, \zeo{-SYT}, and \zeo{-ITV} tend to cluster together, as their distances to all other zeolites is high (Fig. \ref{fig:si:01-analysis-distributions}).
At the same time, other interrupted frameworks produced in different synthesis conditions, such as \zeo{-IRY}, \zeo{-IFU}, or \zeo{-IFT} synthesized with germanium, appear along with other germanosilicates in the synthesis map, not only with other interrupted frameworks.
This ability to traverse the structural space in a continuous way allows drawing non-obvious connections that may be overlooked by building units alone.
For example, the \zeo{MEI} framework lies within the \zeo{SBS}, \zeo{SBT}, \zeo{SAO}, and \zeo{SBE} cluster despite not having common CBUs with any of these zeolites.
Nevertheless, the \zeo{MEI} structure also has large pores and building patterns that could help inform the synthesis of other of these zeolites.
In fact, the recent synthesis of PST-32 (\zeo{SBT}) and PST-2 (\zeo{SBS/SBT}) aluminosilicate zeolites was performed in the presence of sodium \cite{Lee2021SynthesisThermally}, similar to the original inorganic synthesis of \zeo{MEI}, although at lower Si/Al ratios.

\begin{figure}[!h]
    \centering
    \includegraphics[width=0.7\linewidth]{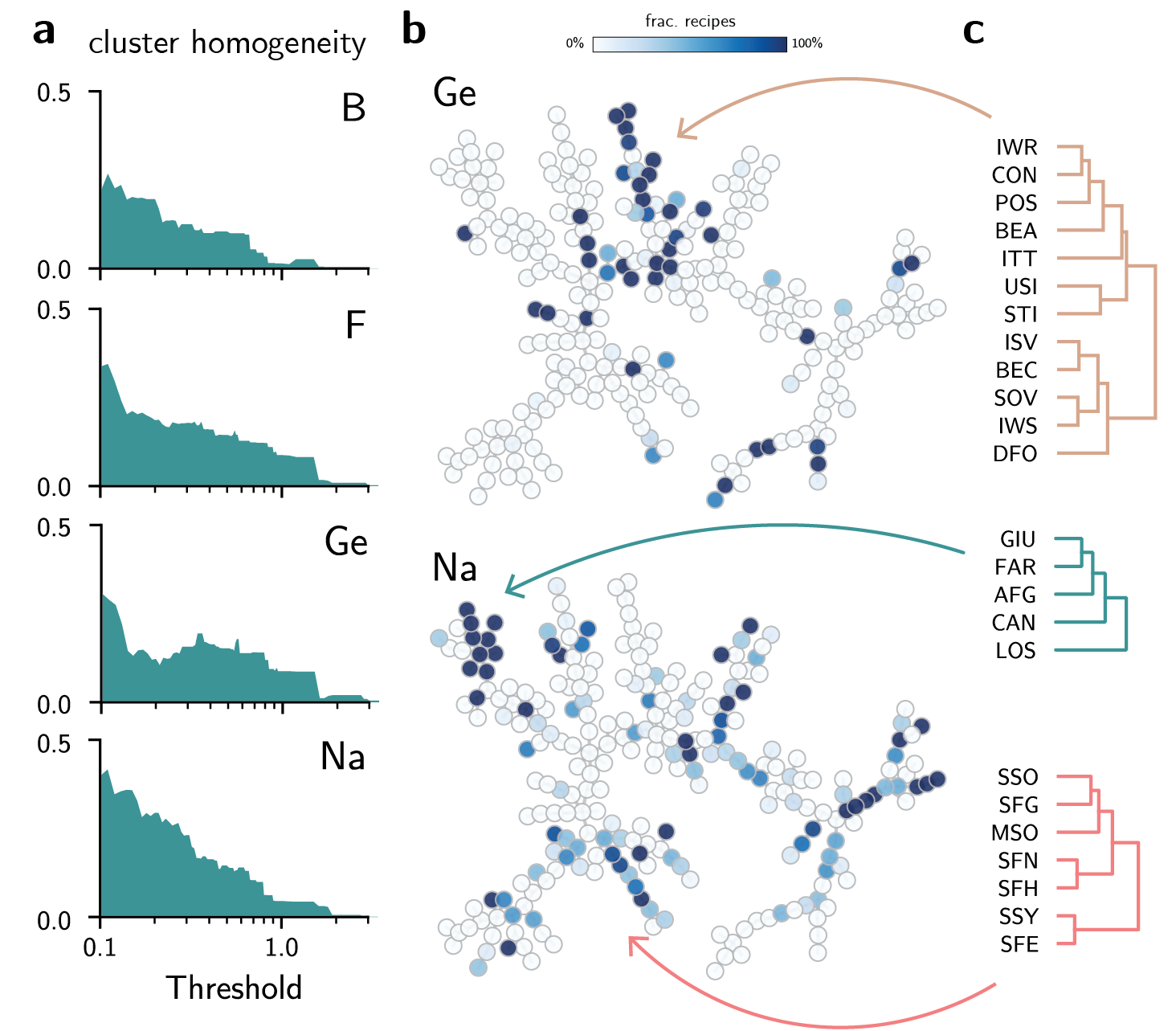}
    \caption{\textbf{Unsupervised learning for inorganic synthesis of known zeolites.} \textbf{a}, Cluster homogeneity of zeolites for selected elements (see also Fig. \ref{fig:si:03-synthesis-homogeneity}). \textbf{b}, Minimum spanning tree of zeolites (Fig. \ref{fig:02-mst}) labeled according to frequency of Ge or Na in the synthesis of each zeolite (see also Fig. \ref{fig:si:03-synthesis-graphs}). Darker (lighter) colors indicate that the inorganic synthesis condition is more (less) frequent in the synthesis of each zeolite. \textbf{c}, Subset of the zeolite dendrogram for selected regions of the minimum spanning tree.}
    \label{fig:03-unsupervised}
\end{figure}

Despite the usefulness of the tree map in connecting zeolites with similar synthesis conditions, the visual analysis cannot determine whether the map consistently provides new insights on the synthesis of zeolites.
To improve this qualitative analysis, we performed a hierarchical clustering of the data to quantify whether the structural distance metric clusters the data according to the literature synthesis conditions (see Methods). 
The dendrogram of the AMD values (Figs. \ref{fig:si:03-dendrogram-full} and \ref{fig:si:03-dendrogram-clusters}) shows how zeolites are related to each other based on distances, thus providing a more quantitative view to the minimum spanning tree of Fig. \ref{fig:02-mst}.
Then, to create labels for synthesis conditions, we started with a dataset of extensive synthesis conditions extracted from the zeolite literature from Jensen et al. \cite{Jensen2021DiscoveringRelationships}.
After augmenting the data with frameworks not typically reported in publications, such as those found as minerals, we analyzed the frequency of occurrence of each synthesis condition for each framework.
Although the initial dataset had information on both organic and inorganic conditions, we disregarded the organic templates when labeling the data, thus assuming that inorganic and organic conditions can, to an extent, be predicted independently of each other.
Furthermore, given the scarcity of data for some synthesis conditions, we focused only on the 14 inorganic conditions that have been used to synthesize at least 10 zeolites, namely Al, B, Be, Ca, Co, F, Ga, Ge, K, Mg, Na, P, Si, and Zn.
Finally, we verify whether flat clusters formed by points with a maximum distance of each other share the same positive labels.
This intuition is quantified by computing the homogeneity between data points given clusters formed by a given distance threshold \cite{Rosenberg2007VMeasure} (see Methods).
If all clusters had only positive labels, their homogeneity would be 1, whereas zero homogeneity indicates perfect mixing of positive and negative labels.
Figure \ref{fig:03-unsupervised}a shows that clusters with at least one positive data point become more homogeneous as the distance threshold decreases.
This supports the qualitative view that structures considered similar according to the AMD values also share similar synthesis conditions more often than not.
On the other hand, as clusters become larger and the increasingly dissimilar structures are grouped together, the homogeneity decreases.
Whereas the distribution of labels for some inorganic agents such as Al, Si, Be, F, or Na exhibit higher homogeneity at low distances (see Fig. \ref{fig:si:03-synthesis-homogeneity}), others such as Co, Mg, or Zn show little predictive power.
This suggests that structural distances computed with the AMD have stronger correlations with certain synthesis conditions than with others.
Although the lack of true negative data points may also influence the computed homogeneity, these results demonstrate quantitatively that some synthesis-structure relationships can be established in zeolites using AMDs.

Additional investigations of the data explain the patterns in homogeneity obtained above.
Figure \ref{fig:03-unsupervised}b shows how the minimum spanning tree can be visualized according to the frequency of certain inorganic conditions in zeolite synthesis (see Fig. \ref{fig:si:03-synthesis-graphs} for complete results).
For example, some frameworks realizable with Ge or Na form their own groups in the tree, as also illustrated by the subclusters in dendrograms (Fig. \ref{fig:03-unsupervised}c).
Indeed, zeolites such as \zeo{BEC}, \zeo{ISV}, \zeo{ITT} or \zeo{IWR} are typical examples of large- and extra-large pore structures synthesized using germanium.
Similarly, denser phases such as \zeo{GIU}, \zeo{FAR}, \zeo{LOS} etc. are often obtained in sodium-mediated syntheses.
For common synthesis conditions such as silicon, trends can be derived from the visualization of silicon-free routes.
The labeled trees from Figs. \ref{fig:02-mst} and \ref{fig:si:03-synthesis-graphs} show that non-silica zeolites are often located in similar regions of the structure space.
Groups formed by zeolites such as \zeo{NAT}, \zeo{EDI}, and \zeo{THO}, or \zeo{AFO}, \zeo{AEL}, \zeo{AHT} show that non-silica zeolites also share structural patterns that may be harder to obtain in silica-based structures.

This unsupervised analysis demonstrates that our distance metric relates zeolites with similar inorganic synthesis conditions without any learnable parameters.
Although zeolite structures contain several outliers and lack true negative data, the structural patterns still provide a strong prior for exploring the synthesis conditions.
In particular, as inorganic synthesis conditions can be inferred by the similarity between crystal structures, they can also help downselect structures for zeolites yet to be realized.

\subsection{Interpretable classifiers for predicting inorganic synthesis conditions}

One disadvantage of the pure unsupervised learning approach is the inefficient utilization of the available labels.
Although similarity between crystal structures is a good indicator of common synthesis conditions, the \textit{dissimilarity} between structures can also provide insights on which structures are less likely to be synthesized with a given composition.
To perform this analysis, we use the labeled data to train supervised learning methods that predict the synthesis conditions of a zeolite given its distances to known frameworks.
Specifically, we trained logistic regression, random forest, and XGBoost classifiers on literature data to predict each class label individually.
However, training models on the literature labels has two caveats: 
(1) the data is often unbalanced, i.e., the number of positive data points is much smaller than the number of negative data points; and 
(2) the negative data is not truly negative, as its lack of literature reporting does not imply that a zeolite cannot be synthesized under the synthesis conditions in analysis. 
To account for these problems, we trained balanced classifiers by subsampling the dataset for each synthesis conditions, thus ensuring that training sets had the same proportion of positive and negative data points, but validation/test sets were allowed to have more negative samples than positive ones.
In that case, because models were tested on different negative splits, they were prevented from memorizing ``negative'' data points as truly negative, as exemplified by the case of \zeo{MEI} zeolite above.
Finally, for each synthesis condition, we performed an extensive hyperparameter optimization for each of the three classifiers, evaluating the models according to their accuracy, precision, recall, F$_1$ score, and areas under the receiving operating characteristic (ROC) and precision-recall (PR) curves.

\begin{figure}[!h]
    \centering
    \includegraphics[width=\linewidth]{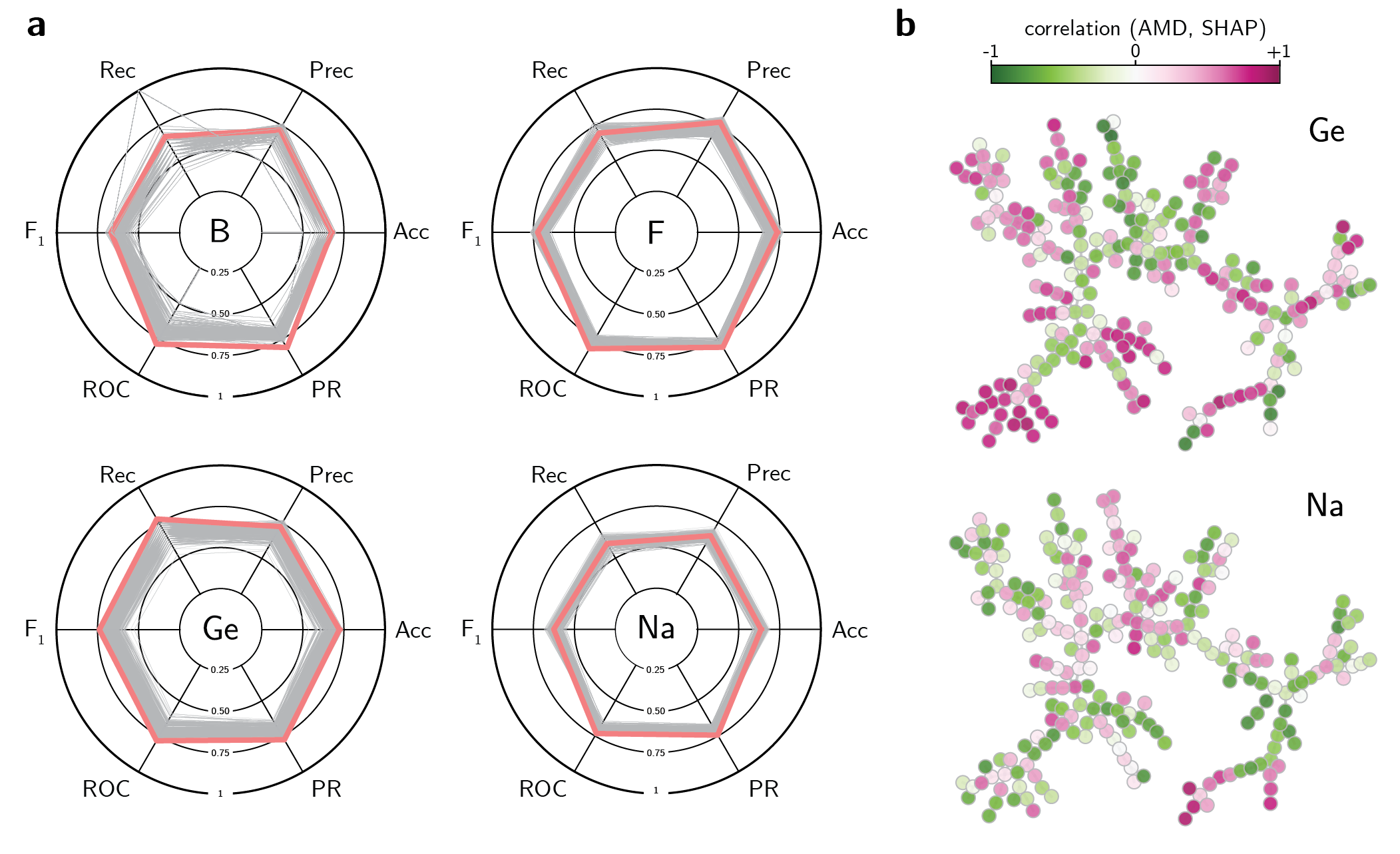}
    \caption{\textbf{Supervised learning of inorganic synthesis in known zeolites.} \textbf{a}, Results of hyperparameter optimization of XGBoost classifiers for selected inorganic conditions. Each thin gray line is one set of parameters for the XGBoost models. The pink line represents the best performance according to the ROC and PR AUC. The figures of merit are: accuracy (Acc), precision (Prec), recall (Rec), F$_1$-score (F$_1$), receiving operating characteristic AUC (ROC), and precision-recall AUC (PR). The figure of merit has value 0 at the center and 1 at the border of the circle. \textbf{b}, Pearson correlation coefficient between AMD and SHAP values. A negative correlation (green) indicates that smaller distances lead to higher SHAP values, and thus contribute to classifying the zeolite with a positive label.}
    \label{fig:04-supervised}
\end{figure}

The results of the hyperparameter search are summarized in Fig. \ref{fig:si:04-classifiers-comparison}.
Whereas no classifier outperforms the other in all tasks, XGBoost models often show the best metrics of ROC and PR areas under the curve (AUC) for a variety of synthesis conditions.
When the results for the XGBoost classifier are visualized according to all metrics at once (Fig. \ref{fig:04-supervised}a), they demonstrate how the best hyperparameters lead to adequate figures of merit based on results from the validation set (see also Fig. \ref{fig:si:04-xgboost-metrics}).
When evaluated against a held-out test set, the model with best set of hyperparameters still exhibits high ROC and PR AUCs for a variety of synthesis conditions (Fig. \ref{fig:si:04-xgboost-auc}).
Nevertheless, this set of hyperparameters is far from being the only one that performs well in these conditions (Fig. \ref{fig:si:04-xgboost-best}).
As discussed in the analysis using unsupervised learning, the ability to correctly label zeolites whose synthesis contains Co or Zn is smaller than other labels, as indicated by the worse performance of all classifiers in labeling these conditions.
However, some synthesis conditions that were not well-predicted by the unsupervised learning method, such as Mg, can now be predicted using XGBoost models, despite its low recall (Fig. \ref{fig:si:04-xgboost-auc}).
These results show that ML  classifiers can predict inorganic synthesis conditions using distances towards known zeolites as features.
This has useful implications, as it bypasses the need to create general representations for zeolites, and instead uses a set of points in the space as references for new synthesis conditions.

To interpret the outcomes of the classifiers, explainability techniques can be used to probe what features most affect a positive (or negative) classification of a zeolite within certain synthesis conditions.
Given that the input features are distances towards known frameworks, a trained classifier decides how to assign a label to an input structure based on its similarity values.
Using the Shapley value method (SHAP), we analyze what distances most affect the classification of a zeolite into a given class.
As each SHAP value indicates how much each feature affects the probability of classifying a framework into a given class, we compute the Pearson correlation coefficient between the actual feature value and the SHAP value for each one of the inorganic synthesis conditions.
This quantifies whether being close to a particular framework (feature) increases or decreases the likelihood of being assigned a positive label.
The results for the interpretability of XGBoost classifiers are shown in Fig. \ref{fig:04-supervised}b (see also Fig. \ref{fig:si:05-classifier-shap} and \ref{fig:si:05-classifier-explainability}).
As the correlation coefficient between AMD and SHAP values are computed on a per-feature (thus per-zeolite) basis, the nodes from the tree map in Fig. \ref{fig:02-mst} are colored according to these coefficients.
In this plot, a negative correlation (in green color) indicates that low distances increase the SHAP value and thus the likelihood of being classified as a positive label.
Conversely, a positive correlation (in pink color) with a feature indicates that a given zeolite is more likely to be synthesized with a given synthesis condition if it is distant from the featurizing structure.
The results not only support the observations highlighted by the unsupervised learning methods, but also complement them with new insights.
For instance, zeolites synthesized with Ca and K have a wide overlap of positive and negative correlations (Fig. \ref{fig:si:05-classifier-explainability}), possibly due to the clustering of minerals in the tree map.
There is also an overlap between boron-containing zeolites and germanium-containing zeolites, especially in the \zeo{ISV} branch.
This result could be interesting if validated in practice, especially as the use of boron may enable the removal of Ge from the synthesis of certain zeolites, such as \zeo{BEC}.
The central branch characterized by Ge-containing zeolites (such as \zeo{BEC}, \zeo{ISV}, \zeo{IRR}, \zeo{ITT}, see Fig. \ref{fig:02-mst}) also have features that correlate with F or Mg, but not Al or Ca.
On the other hand, Be-containing zeolites are often complementary to Si-containing zeolites, as the former are found only in specific clusters or outliers in the tree map.
This further supports the fact that the classifiers are able to obtain correlations beyond existing heuristics, thus providing data-driven ways to guide inorganic synthesis in zeolites.

\subsection{Proposing inorganic synthesis conditions for hypothetical zeolites}

Given that structural similarity is correlated to inorganic synthesis in zeolites and that supervised learning methods are able to predict synthesis using only distances as inputs, we can use the models developed in this work to propose inorganic synthesis conditions for hypothetical zeolites.
This approach complements previous work on the design of OSDAs for frameworks, thus enabling inorganic synthesis conditions to be predicted \textit{in silico} whenever a new framework is proposed.
To do that, we used the dataset of 331,171 hypothetical zeolites proposed by Pophale et al. \cite{Pophale2011DatabaseNew}, known as the ``Deem dataset.''
As structural features and densities from the hypothetical zeolites optimized with force fields may deviate from the experimental ones, we used the IZA and hypothetical zeolites from Erlebach et al. \cite{Erlebach2022AccurateLargescale}, which employed a neural network force field trained at the SCAN level of density functional theory calculations to reoptimize the hypothetical zeolites.
Then, by comparing the hypothetical structures against all known zeolites, we created a distance matrix that is used as input for the unsupervised and supervised learning methods shown in the previous section.
As in the case of known zeolites, AMDs are correlated with differences of density, but are not solely determined by them (Fig. \ref{fig:si:06-hyp-analysis}).
Using AMDs, a low-dimensional map can be created for all hypothetical structures, thus providing an intuitive way to visualize the space of structures.
Figure \ref{fig:si:06-hyp-umap} shows a 2D projection of the distribution of hypothetical zeolites based on their distance matrix using UMAP.
This plot shows that distance features are able to sort the space of zeolites according to energy and density despite not using this information as explicit inputs.
The visualization also illustrates that most hypothetical frameworks do not have neighboring known structures.
While 105 of all known zeolites have at least one other known zeolite up to 0.1 \AA~ away (45\% of structures, see Fig. \ref{fig:si:03-dendrogram-clusters}), only about 36,112 of the 331,171 hypothetical structures have at least one known zeolite as neighbor when the same distance threshold is used (11\% of zeolites in the dataset).
This illustrates how the space of enumerated zeolites is often populated with structures far from known structural patterns of zeolites, as also demonstrated by previous studies (see also Fig. \ref{fig:si:06-hyp-density}).

As demonstrated in this work, zeolites in the neighborhood of known frameworks are likely to share similar synthesis conditions as those known structures.
Thus, downselecting frameworks for given synthesis conditions can benefit from the unsupervised and supervised methods developed here.
This approach can be used in combination with previous ``synthesizability metrics'' of zeolites, such as local interatomic distances \cite{Li2013CriteriaZeolite} or other data-driven predictions \cite{Helfrecht2022RankingSynthesizability}.
However, we chose to evaluate them independently, as these synthesizability predictions do not take into account that certain known frameworks may be considered ``unfeasible'' depending on the synthesis conditions \cite{Mazur2016SynthesisUnfeasible,Li2019NecessityHeteroatoms}.
For instance, structures containing three-connected rings, such as those with building units \textit{lov} or \textit{vsv}, could be ranked as ``unsynthesizable,'' despite being achieved with beryllium or borogermanate conditions.
Thus, to propose synthesis conditions for zeolites, we evaluated all hypothetical frameworks for all synthesis conditions using an ensemble of 100 binary classifiers per inorganic condition (see Methods).
As each classifier is trained on different negative data splits, the resulting classification varies for each model, allowing us to assess the degree of agreement between the models.
By taking the average of the predictions, we obtain the probability of synthesizing the zeolite with that synthesis condition.

\begin{figure}[!h]
    \centering
    \includegraphics[width=\linewidth]{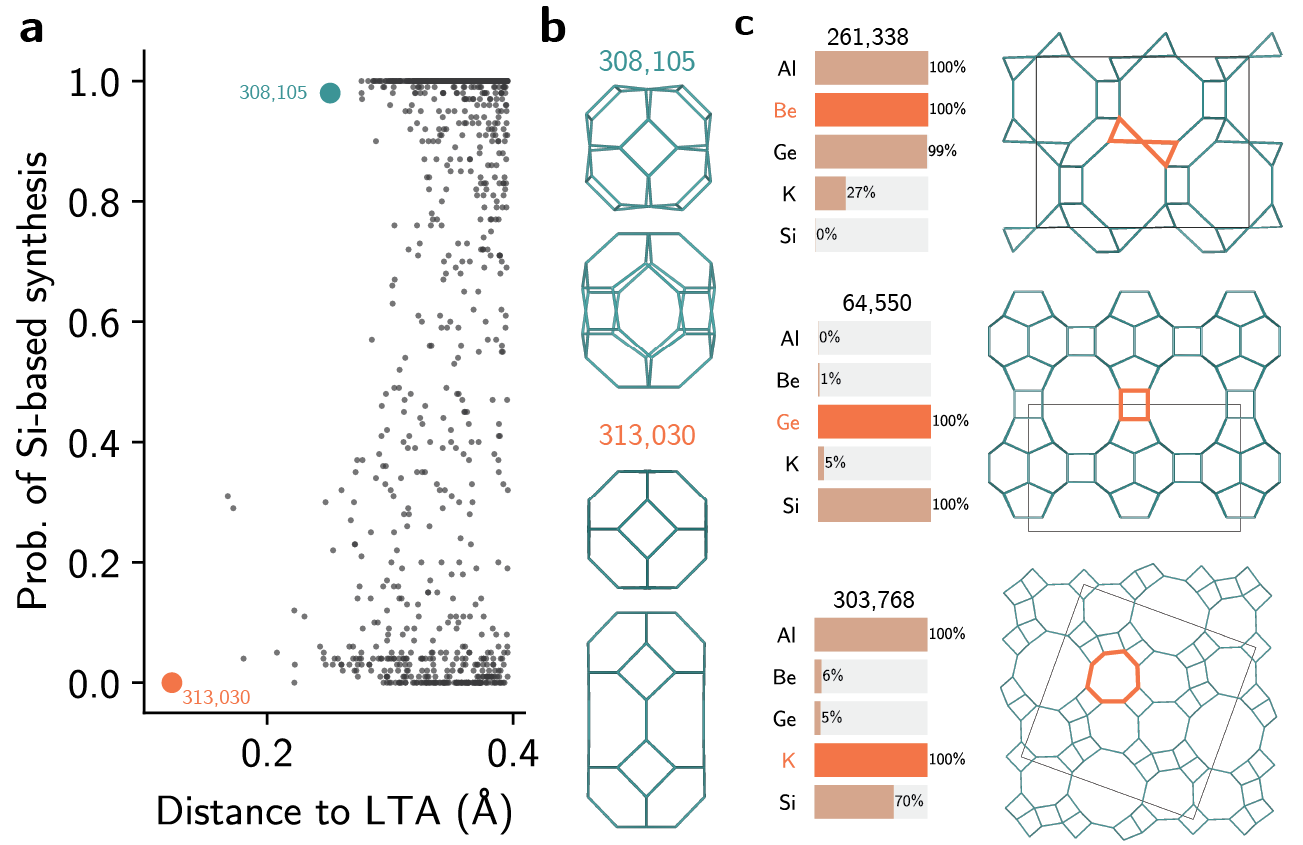}
    \caption{\textbf{Selection of hypothetical zeolites using AMD values and inorganic conditions.} \textbf{a}, Selection of \zeo{LTA}-like zeolites according to predicted Si-based recipes. Only the 1000 closest points to \zeo{LTA} are shown in this graph. \textbf{b}, Unusual cages that distinguish the two hypothetical structures \#308,105 and \#313,030 and determine their synthesis to be more/less likely to be successful under Si conditions, respectively. \textbf{c}, Three examples of hypothetical zeolites selected based on the predicted synthesis conditions. Only five (Al, Be, Ge, K, Si) of the 14 synthesis conditions are shown.}
    \label{fig:06-hyp}
\end{figure}

Figure \ref{fig:06-hyp}a depicts the distribution of hypothetical zeolites with Si-based recipes in the neighborhood of \zeo{LTA} zeolite.
As all distances between known and hypothetical zeolites had been already computed, we can use both the distances and the class probabilities as criteria for navigating the space of hypothetical structures.
This navigation using reference materials instead of features facilitates the selection process and can also inform their synthesis.
For example, Figs. \ref{fig:06-hyp}a,b illustrate two different hypothetical zeolites in the neighborhood of \zeo{LTA}.
Although both have low distance towards \zeo{LTA} (compare with dendrogram in Fig. \ref{fig:si:03-dendrogram-clusters}), structure \#308,105 is predicted to be more likely to be synthesized as a silicate than \#313,030.
Both contain the \textit{lta} and \textit{sod} cages characteristic of the \zeo{LTA} zeolite, but differ by the presence of a second cage similar to \textit{sod}, shown in Fig. \ref{fig:06-hyp}b.
Whereas this new building unit resembles an expanded \textit{sod} cage with distorted six-membered rings in \#308,105, hypothetical framework \#313,030 shows a new cage, formed by the merging of two \textit{sod} cages, not seen in known zeolites.
This increased distance towards known structural patterns drives the prediction of feasible synthesis using Si as unlikely, even when the distance towards the \zeo{LTA} zeolite is lower.
This example shows how the combination of AMD values and classifier predictions facilitates the exploration of hypothetical zeolites using reference structures.

Beyond exploration of the zeolite space, the models also uncover existing and new synthesis-structure relationships.
Figure \ref{fig:06-hyp}c shows three examples of hypothetical frameworks predicted to be synthesized using three different elements: Be, Ge, and K.
To obtain these frameworks, we filtered only frameworks within densities of 14 and 17 T/1000 \AA$^3$ that are predicted to have 100\% probability of synthesis with the given element.
Then, we ranked the frameworks by their relative energy.
Despite not using explicit labels on the CBUs, the supervised learning models recovered the known heuristics of building units and inorganic synthesis conditions.
For instance, framework \#261,338, predicted to be synthesized in presence of Be, is formed mostly by \textit{lov} building units, as found in other Be-zeolites such as \zeo{RSN}, \zeo{LOV}, or \zeo{NAB}.
This same framework is predicted to be unlikely as a silicate, possibly following the trends seen in \zeo{JSR} or \zeo{NPT} structures.
Hypothetical zeolite \#64,550, predicted to be synthesized with germanium, also shows features similar to known ones.
In addition to its three-dimensional pore structure, with $12 \times 12 \times 10$ intersecting pores, the structure shows the \textit{d4r} CBU typical of other structurally similar germanosilicates, such as \zeo{POS} or \zeo{UOV}, but with 7 symmetrically inequivalent T sites.
Finally, one unrealized framework predicted to be synthesized with potassium is structure \#303,768.
Although this hypothetical structure does not exhibit typical CBUs, the local structures similar to \textit{d8r} CBUs are predicted to be favored by K, in analogy with similar relationships in known zeolites.
This demonstrates how data-driven models can not only recover known relationships between CBUs and inorganic conditions, but also propose new synthesis-structure relationships in zeolites based on distance patterns towards known structures.
When used to analyze the entire space of hypothetical frameworks, the models show that the distribution of predicted inorganic synthesis conditions is uneven across the space of zeolites (Fig. \ref{fig:si:06-hyp-inorganics}).
For instance, whereas about 27\% of all known zeolites can be synthesized with germanium, according to the literature dataset we used in this work, only 8\% of the hypothetical zeolites have probability of being synthesizable under Ge conditions of at least 80\%.
Similarly, the space of hypothetical structures is surprisingly lacking in structures predicted to be synthesizable with sodium.
While about 45\% of all known structures have at least one sodium-based synthesis, 17\% of hypothetical structures are predicted to be realizable with Na given the 80\% threshold probability.
As most enumerated datasets are often created without considering synthesis conditions \cite{Argente2023ComputerGeneration}, comprehensive enumerations may introduce biases in structures that do not reflect the space of zeolite synthesis typically considered in practice.
Thus, in combination with template design \cite{Schwalbe-Koda2021PrioriControl} and property screening \cite{Lin2012SilicoScreening,Hewitt2022MachineLearning}, our methods to predict inorganic synthesis conditions in zeolites may help in synthesizing unrealized frameworks with targeted properties.

\section{Conclusions}

Mapping the space of inorganic conditions in materials synthesis is an outstanding challenge due to the complexity of chemical interactions during synthesis.
In the case of zeolites, synthesis conditions are known to affect structural patterns in the materials, but finding correlations between structural patterns and inorganic syntheses often relies on heuristics.
In this work, we used unsupervised and supervised learning methods to propose inorganic synthesis conditions for zeolite synthesis.
In particular, we showed how a mathematically strong distance metric between crystals can predict inorganic synthesis conditions in zeolites.
This enables structural comparisons beyond human-crafted labels of building units or pore sizes/topologies.
Clustering techniques demonstrate that our metric consistently recalls the inorganic synthesis conditions from literature datasets, thus providing predictive power even in the absence of labels.
Then, we show that ML classifiers can predict synthesis conditions of a given framework based on its distribution of distances towards known structures.
The classifiers were used to predict 14 different synthesis conditions for known and unrealized zeolites.
When explaining the predictions, we showed how ML classifiers analyze synthesis conditions also from the dissimilarity between crystals, as well as from the similarity.
The results from the explainability analysis reveals overlaps in inorganic synthesis conditions, such as boron and germanium, as well as complementary relationships, such as silicon and beryllium.
Finally, we showcased how our methods can be used to predict inorganic synthesis conditions for hypothetical zeolites, facilitating the downselection of new structures for experimental attempts.
This combination of data-driven methods can create a pathway for full \textit{in silico} prediction of zeolite synthesis beyond the design of organic templates.

\section{Methods}

\subsection{Pointwise Distance Distributions and Average Minimum Distances}

Any periodic crystal structure is modeled as a periodic set $S$ of atomic centers considered as zero-sized points, with atomic types as optional labels.
Any linear basis of vectors $v_1,v_2,v_3$ in 3-dimensional space generates a lattice $\Lambda=\{c_1v_1+c_2v_2+c_3v_3 \mid c_i \text{ are integers}\}$ and unit cell $U=\{t_1v_1+t_2v_2+t_3v_3 \mid 0\leq t_i<1\}$.
Any finite motif of points $M\subset U$ defines the periodic point set $S=\{p+ v \mid p\in M, v\in\Lambda\}$.
This conventional representation of a periodic crystal $S$ by a unit cell and a motif is ambiguous because infinitely many different pairs (cell, motif) generate periodic sets that are equivalent up to rigid motion (a composition of translations and rotations).
Fixing any reduced cells leads to unavoidable discontinuities \cite{kurlin2022mathematics} even for 2-dimensional lattices.

The ambiguity of crystal representations was theoretically resolved for all periodic point sets in any dimension by the complete isoset \cite{anosova2021isometry} invariant.
We define below the computationally faster Pointwise Distance Distribution (PDD) invariant, which distinguished all (more than 670,000) periodic crystals in the Cambridge Structural Database (CSD) through more than 200 billion pairwise comparisons within two days on a typical desktop computer.

\begin{figure}[!h]
    \centering
    \includegraphics[width=\linewidth]{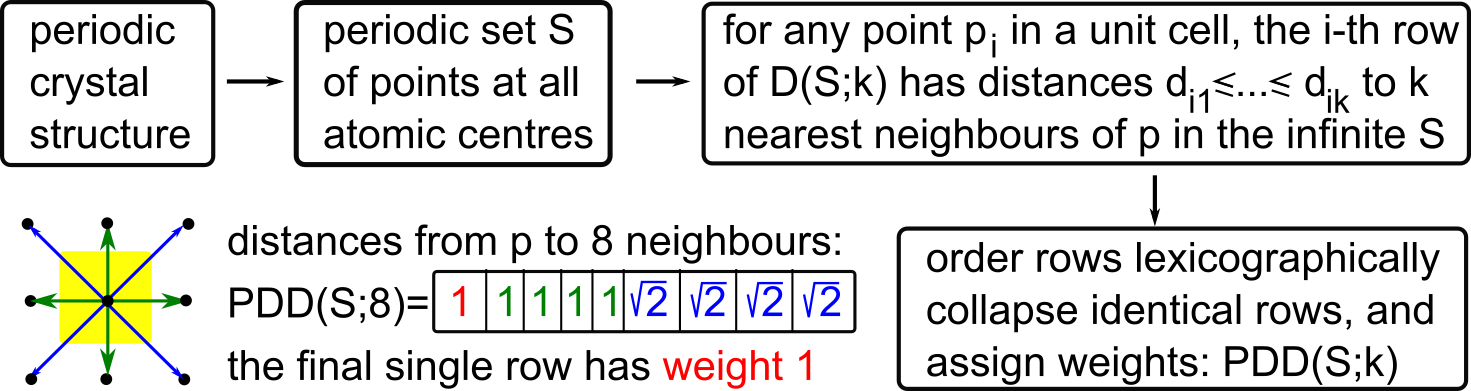}
    \caption{Computational pipeline of PDD is illustrated for a 2-dimensional square lattice.}
    \label{fig:07-PDD}
\end{figure}

Fix a number $k\geq 1$ of atomic neighbors.
Our experiments on zeolites and the CSD used $k=100$.
Let $S$ be a periodic set with a motif $M$ of points $p_1,\dots,p_m$.
For each point $p_i$, write down the sequence of increasing distances $d_{i1}\leq\cdots\leq d_{ik}$ to its $k$ nearest neighbors in the full infinite set $S$ without considering any extended cell or cut-off radius.
Collect these sequences of distances into an $m\times k$ matrix and lexicographically order the rows.
If any $l$ of the rows coincide (usually due to extra symmetries), collapse them into a single row with the weight $l/m$ and put these weights into an additional first column (unique rows have weight $1/m$).
The resulting $m \times (k + 1)$ matrix $\mathrm{PDD}(S;k)$ is called the \emph{Pointwise Distance Distribution}, a statistical distribution of rows with weights describing each point's environment.
As an example, Fig.~\ref{fig:07-PDD} shows the computation for a point in the square lattice $S$ whose first $k=8$ neighbours have distances $1,1,1,1$ (in green) and $\sqrt{2},\sqrt{2},\sqrt{2},\sqrt{2}$ (in blue).

The \emph{Average Minimum Distance} $\mathrm{AMD}(S;k)$ is the vector obtained by taking the weighted average of the last $k$ columns in $\mathrm{PDD}(S;k)$, so $\mathrm{AMD}$ is a single vector of $k$ average distances.
To compare two AMD vectors of the same length, our experiments used the $L_\infty$ metric equal to the maximum absolute difference of corresponding coordinates.
For a metric on PDDs, we use the Earth Mover's Distance (EMD)\cite{hargreaves2020earth} with the $L_\infty$ metric on rows. 
If any point of $S$ is perturbed in its $\varepsilon$-neighborhood, then $\mathrm{PDD}(S;k)$ changes by at most $2\varepsilon$ in the EMD metric. 
If a periodic set $S$ is generic, which is achievable by almost any perturbation of atoms, then $S$ can be reconstructed from the number $m$ of motif points, a (basis of a) lattice $\Lambda$ and $\mathrm{PDD}(S;k)$ with a known upper bound on $k$.
For the details on these results, see Definition C5 and proofs of Theorems 4.3 and 4.4 in the extended version of Ref. \citenum{widdowson2021pointwise}.

\subsection{Zeolite structures data}

The dataset of 253 known zeolite structures used in the unsupervised learning method was obtained from the International Zeolite Association (IZA) database \cite{Baerlocher2023DatabaseZeolite}.
The dataset of hypothetical frameworks used in this work was developed by Pophale et al. \cite{Pophale2011DatabaseNew}, and re-optimized using a neural network force field trained on DFT-SCAN data by Erlebach et al. \cite{Erlebach2022AccurateLargescale}.
Because not all of the 253 known zeolites used previously were optimized by Erlebach et al., we used their subset of 236 known frameworks when computing distance matrices from the hypothetical frameworks and the known frameworks.

Following the notation from the IZA, known zeolites are named in this manuscript according to their three-letter code in bold typeface.
Known CBUs are represented with their three-letter code in lowercase and italic typeface.

\subsection{Literature data}

Literature data was obtained from public datasets of zeolite synthesis conditions from Jensen et al. \cite{Jensen2021DiscoveringRelationships}, which has been validated by Schwalbe-Koda et al. for the computational design of OSDAs \cite{Schwalbe-Koda2021PrioriControl}.
Whereas in those works only zeolite-OSDA pairs were considered, in this work only relationships between zeolites and non-organic synthesis conditions are analyzed.
As the dataset was produced by collecting literature data from over 60 years of studies in synthetic zeolites, several natural frameworks were omitted from the table, as well as newer structures not captured at the time of this study.
To address this problem, we manually inserted new rows on the table with the composition of the missing structures.
The compositions were obtained with manual verification of the synthesis conditions in articles describing the mineral/synthetic zeolite, as also shown in the online IZA database.
The resulting, cleaned data used in this study is available for download (see Code and Data Availability).

In the literature analysis, a zeolite is classified as having a certain synthesis condition when at least 25\% of its synthesis recipes exhibit that condition (excluding OSDAs).
This label is used as a categorical variable when performing the classification task.

\subsection{Unsupervised learning}

A \textbf{minimum spanning tree} between zeolites was constructed by first creating a fully connected, undirected graph with weighted edges, where weights correspond to the distances between two structures.
The tree was then obtained using NetworkX's (v. 2.5) \cite{Hagberg2008NetworkX} minimum spanning tree algorithm, which minimizes the total length of the tree.

The \textbf{dendrogram} of known zeolites was produced by creating a linkage matrix from the distance matrix using the Ward algorithm as implemented in SciPy (v. 1.10.0) \cite{Virtanen2020SciPy}.
The resulting clusters in Fig. \ref{fig:si:03-dendrogram-clusters} were obtained by forming flat clusters with maximum AMD distance of a given threshold.

The \textbf{homogeneity} of the clustering was computed by calculating the Shannon entropy of flat clusters created with a given threshold \cite{Rosenberg2007VMeasure}, as implemented in scikit-learn (v. 1.2.0) \cite{Pedregosa2011ScikitLearn}.
As the literature dataset is not balanced and lack true negative points, the homogeneity was only computed for clusters containing at least one positive data point.
This ensures that a large homogeneity corresponds to recall of positive data points, which prevents biasing this metric in imbalanced datasets.

\textbf{Dimensionality reduction} was performed using UMAP \cite{McInnes2018UMAP}, as implemented in the \texttt{umap-learn} package in Python (v. 0.5.3).
The 2D UMAP plot was produced by comparing hypothetical frameworks using the cosine distance of their normalized distances to IZA structures, and using 10 neighbors as parameter.

\subsection{Supervised learning}

\textbf{Classification} of inorganic synthesis conditions was performed by training separate classifiers for each synthesis condition.
The features used during training were the distances towards the 253 known frameworks, as computed with the AMD method described above.
To obtain a statistically meaningful result, only elements used to synthesize at least 10 zeolites were considered.
In particular, 14 inorganic conditions are considered: Al, B, Be, Ca, Co, F, Ga, Ge, K, Mg, Na, P, Si, and Zn.

Train-validation-test sets were created starting with a 60-20-20 ratio, respectively, then subsampling the training set to have an equal number of points with positive and negative labels.
Although techniques such as reweighting or resampling could have been employed to obtained balanced training sets,  removing data points is a simple approach that prevents classifiers from treating negative data as ``true negative'', resembling positive-unlabeled learning strategies.

\textbf{Hyperparameter optimization} of synthesis classifiers was performed using a grid-search method over relevant spaces of hyperparameters for logistic regression, random forest, and XGBoost \cite{Chen2016XGBoost} methods.
The full range of hyperparameters investigated in this hyperparameter search is shown in Tables \ref{tab:si:logistic}, \ref{tab:si:rf} and \ref{tab:si:xgb}, following the notation in the \texttt{scikit-learn} (v. 1.2.0) and \texttt{xgboost} (v. 1.7.5) Python packages.
Model performances were compared using the same dataset splits, and the best model is selected according to its validation performance.
The results on the paper showcase the performance on held-out test data.

One of the best models to classify synthesis conditions of zeolites was the \textbf{XGBoost model} with the following hyperparameters: \texttt{colsample\_bytree = 0.5, learning\_rate = 0.1, max\_depth = 6, min\_child\_weight = 1, n\_estimators = 200, subsample = 0.5}.
This model and set of hyperparameters showed good performance across a range of inorganic synthesis conditions, as evaluated by the accuracy, precision, recall, F$_1$ score, area under the receiving operator characteristic curve (ROC AUC), and area under the precision-recall curve (PR AUC).
In particular, the best model was selected to maximize the ROC AUC and PR AUC for the balanced classifiers.
As a comparison, the performance metrics and their baselines of unbalanced classifiers --- thus trained on dataset splits with an uneven number of positive/negative labels --- are shown in Fig. \ref{fig:si:04-xgboost-auc}.

Explainability of the classifiers was computed using the Shapley value method (SHAP) \cite{Lundberg2017UnifiedApproach} under the TreeExplainer framework \cite{Lundberg2020LocalExplanations}, as implemented in the \texttt{shap} Python package (v. 0.41.0). The interventional feature perturbation method was used without limit for the tree explainer.
Then, correlations between the SHAP values and the distance features were computed for each synthesis condition.
To ensure that the correlations are not artifacts of particular train splits, we report the average correlation obtained from an ensemble of 100 XGBoost classifiers trained on splits with different negative data points.



\begin{acknowledgement}
    This work was performed under the auspices of the U.S. Department of Energy by Lawrence Livermore National Laboratory (LLNL) under Contract DE-AC52-07NA27344. D.S.-K. acknowledges funding from the Laboratory Directed Research and Development (LDRD) Program at LLNL under project tracking code 22-ERD-055, and under the Postdoctoral Development Program. T.A.P. acknowledges funding from the Center for Enhanced Nanofluidic Transport (CENT), an Energy Frontier Research Center funded by the U.S. Department of Energy, Office of Science, Basic Energy Sciences under Award DE-SC0019112. D.E.W. and V.A.K. acknowledge funding from the EPSRC (EP/R018472/1, EP/X018474/1) and Royal Academy of Engineering (IF2122/186) in the UK. 

Manuscript released as \texttt{LLNL-JRNL-851183}.
\end{acknowledgement}

\section*{Data and Code Availability}

The code to compute PDDs and AMDs for arbitrary crystal structures is available on GitHub at \url{https://github.com/dwiddow/average-minimum-distance}.
The code and data required to reproduce all results and figures in this manuscript will be available after peer-review.

\providecommand{\latin}[1]{#1}
\makeatletter
\providecommand{\doi}
  {\begingroup\let\do\@makeother\dospecials
  \catcode`\{=1 \catcode`\}=2 \doi@aux}
\providecommand{\doi@aux}[1]{\endgroup\texttt{#1}}
\makeatother
\providecommand*\mcitethebibliography{\thebibliography}
\csname @ifundefined\endcsname{endmcitethebibliography}
  {\let\endmcitethebibliography\endthebibliography}{}

\clearpage

\appendix

\begin{spacing}{2.5}
\begin{center}
    {\fontfamily{lmss}\fontsize{18}{18}\selectfont\textbf{Supporting Information for: Inorganic synthesis-structure maps in zeolites with machine learning and crystallographic distances}}\\[12pt]
    {\fontfamily{lmss}\fontsize{14}{14}\selectfont Daniel Schwalbe-Koda$^{*,\dag}$, Daniel E. Widdowson,$^{\ddag}$ Tuan Anh Pham,$^{\dag}$ and Vitaliy Kurlin$^{*,\ddag}$}\\[12pt]
    \fontsize{12}{12}\selectfont $^\dag$\textit{Lawrence Livermore National Laboratory, Livermore, CA, United States}\\
    $^\ddag$\textit{University of Liverpool, Liverpool, United Kingdom}\\
    {\fontfamily{lmss}\fontsize{12}{12}\selectfont E-mail: dskoda@llnl.gov; vitaliy.kurlin@gmail.com}
\end{center}
\end{spacing}

\renewcommand{\thefigure}{S\arabic{figure}}
\setcounter{figure}{0}
\renewcommand{\thetable}{S\arabic{table}}
\setcounter{table}{0}

\pagenumbering{arabic}
\setcounter{page}{1}

\section{Supporting Figures}

\begin{figure}[!h]
    \centering
    \includegraphics[width=\linewidth]{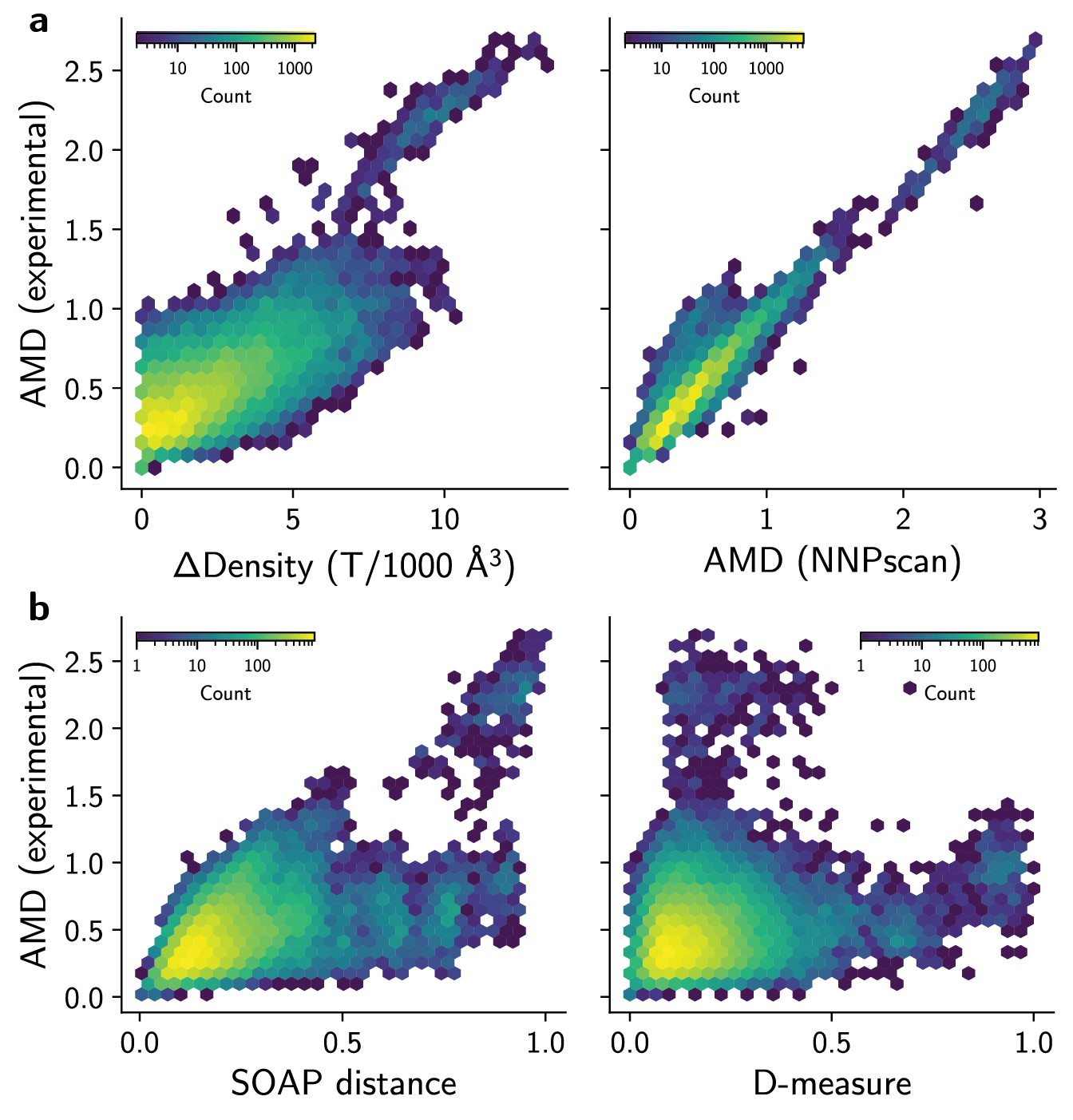}
    \caption{Correlations of AMD values between zeolites with experimental lattice parameters from the IZA database and \textbf{a}, density differences, \textbf{b}, AMD values for zeolites optimized with NNPscan, \textbf{c}, SOAP distance, and \textbf{d}, the graph distance D-measure. The values from \textbf{c, d} were retrieved from Schwalbe-Koda et al. (Ref. \citenum{Schwalbe-Koda2019GraphSimilarity} from the main text)}
    \label{fig:si:01-analysis-distances}
\end{figure}

\begin{figure}[!h]
    \centering
    \includegraphics[width=\linewidth]{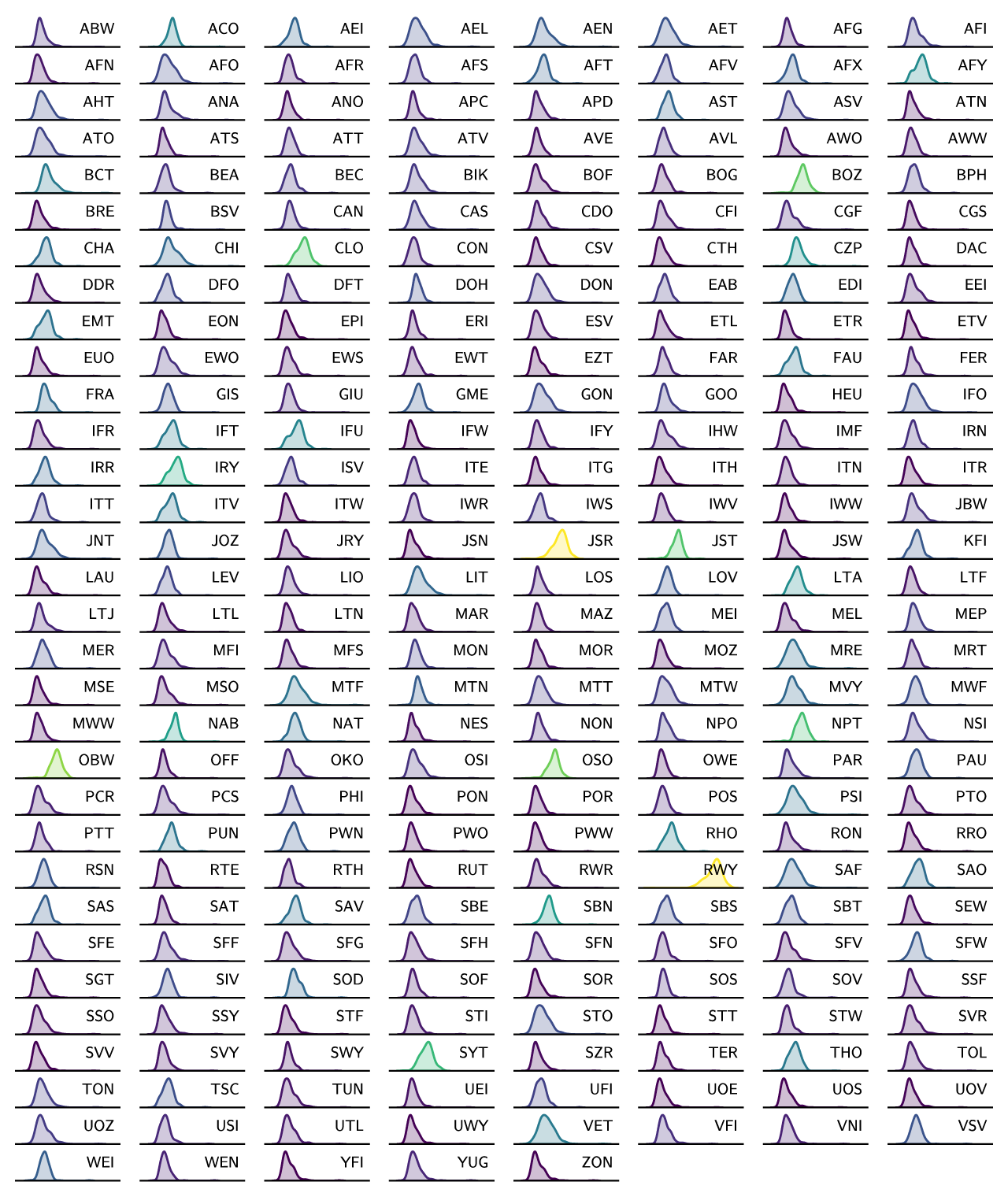}
    \caption{Distributions of AMD distances for each zeolite against all other known zeolites. Brighter colors indicate a higher median. All subfigures share the same x axis.}
    \label{fig:si:01-analysis-distributions}
\end{figure}

\begin{figure}[!h]
    \centering
    \includegraphics[width=\linewidth]{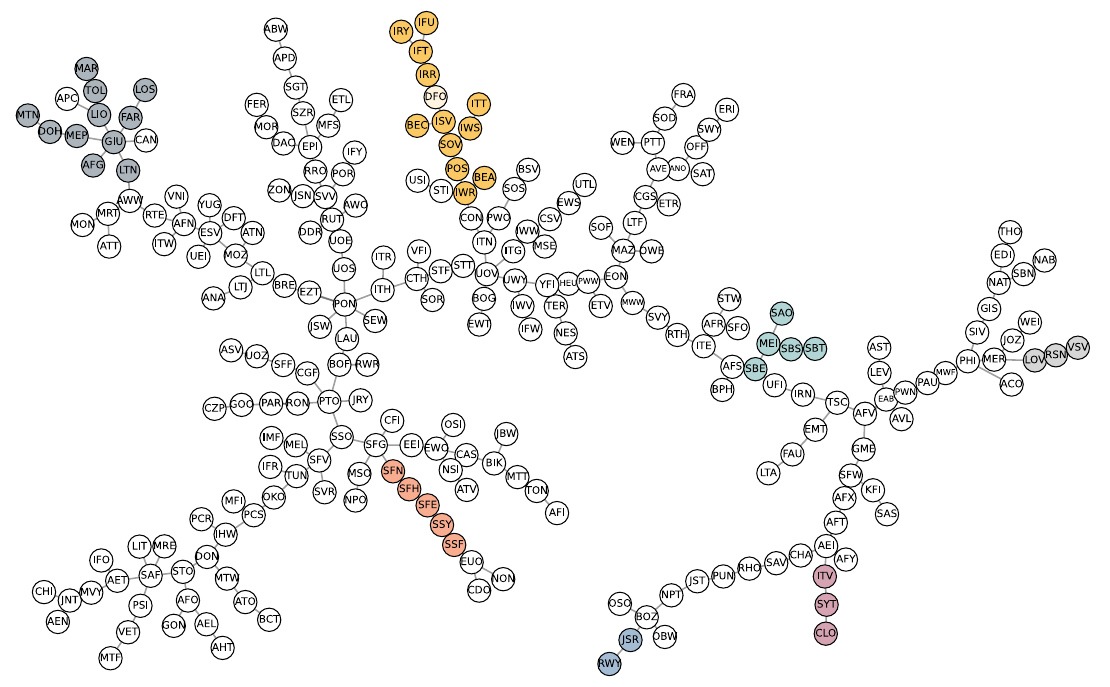}
    \caption{Minimum spanning tree of known zeolites with colored labels for a visual guide. Although several clusters could be highlighted, only some of those discussed in the main text are shown here.}
    \label{fig:si:02-mst-guide}
\end{figure}

\begin{figure}[!h]
    \centering
    \includegraphics[width=\linewidth]{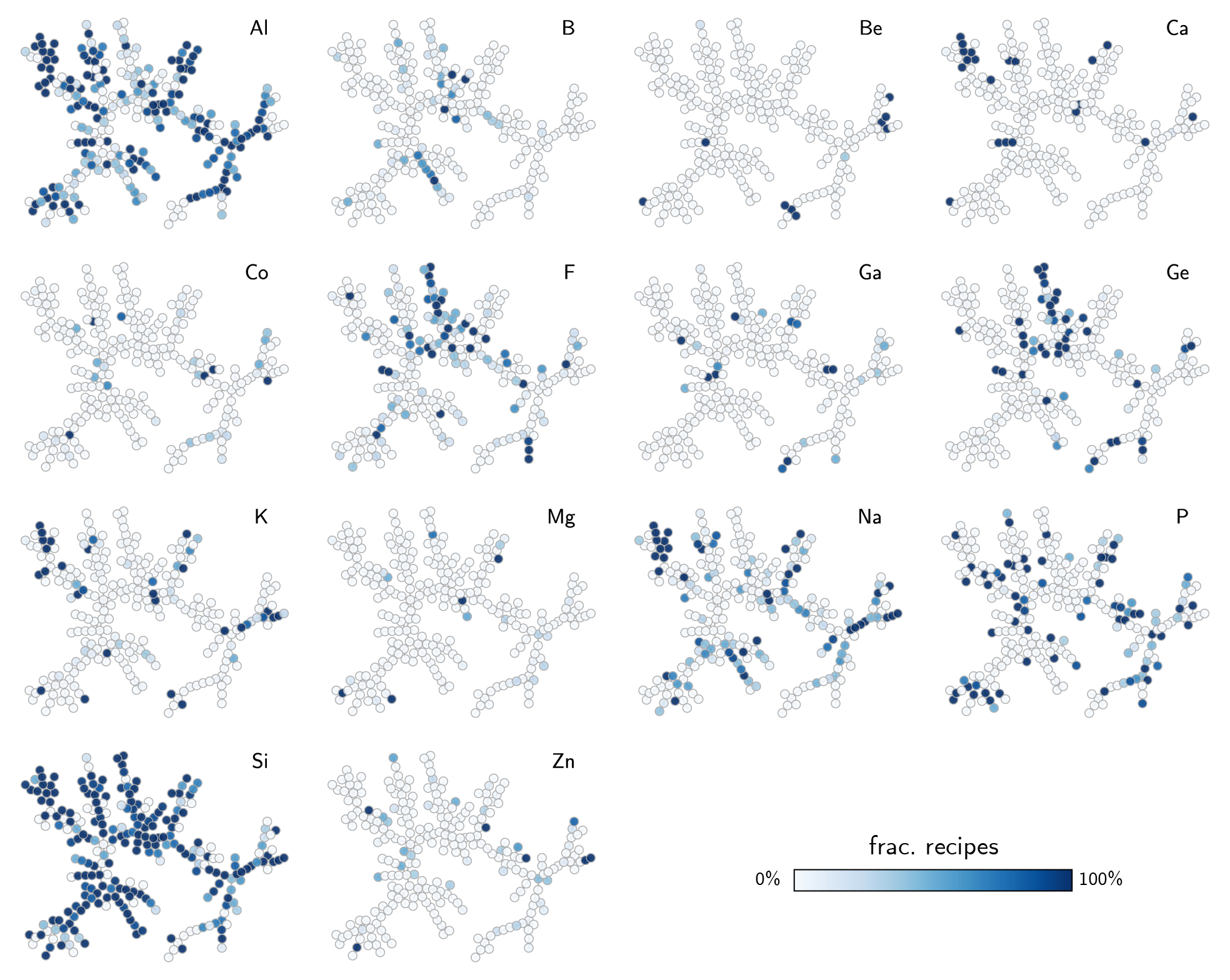}
    \caption{Minimum spanning tree of known zeolites labeled according to the fraction of recipes per element.}
    \label{fig:si:03-synthesis-graphs}
\end{figure}

\begin{figure}[!h]
    \centering
    \includegraphics[width=0.7\linewidth]{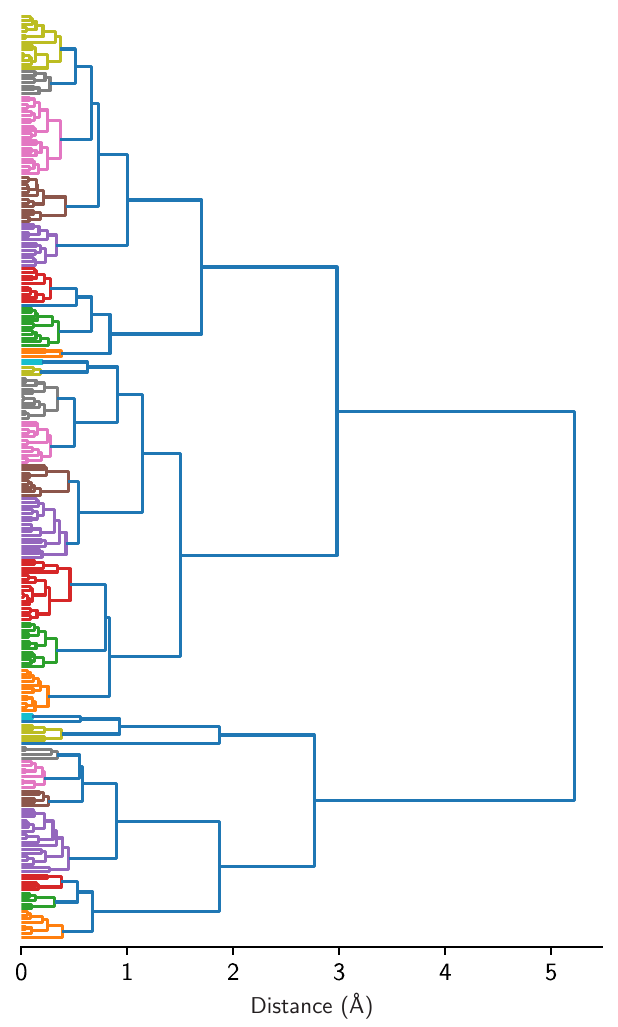}
    \caption{Complete dendrogram of zeolites created using the AMD values as features.}
    \label{fig:si:03-dendrogram-full}
\end{figure}

\begin{figure}[!h]
    \centering
    \includegraphics[width=0.9\linewidth]{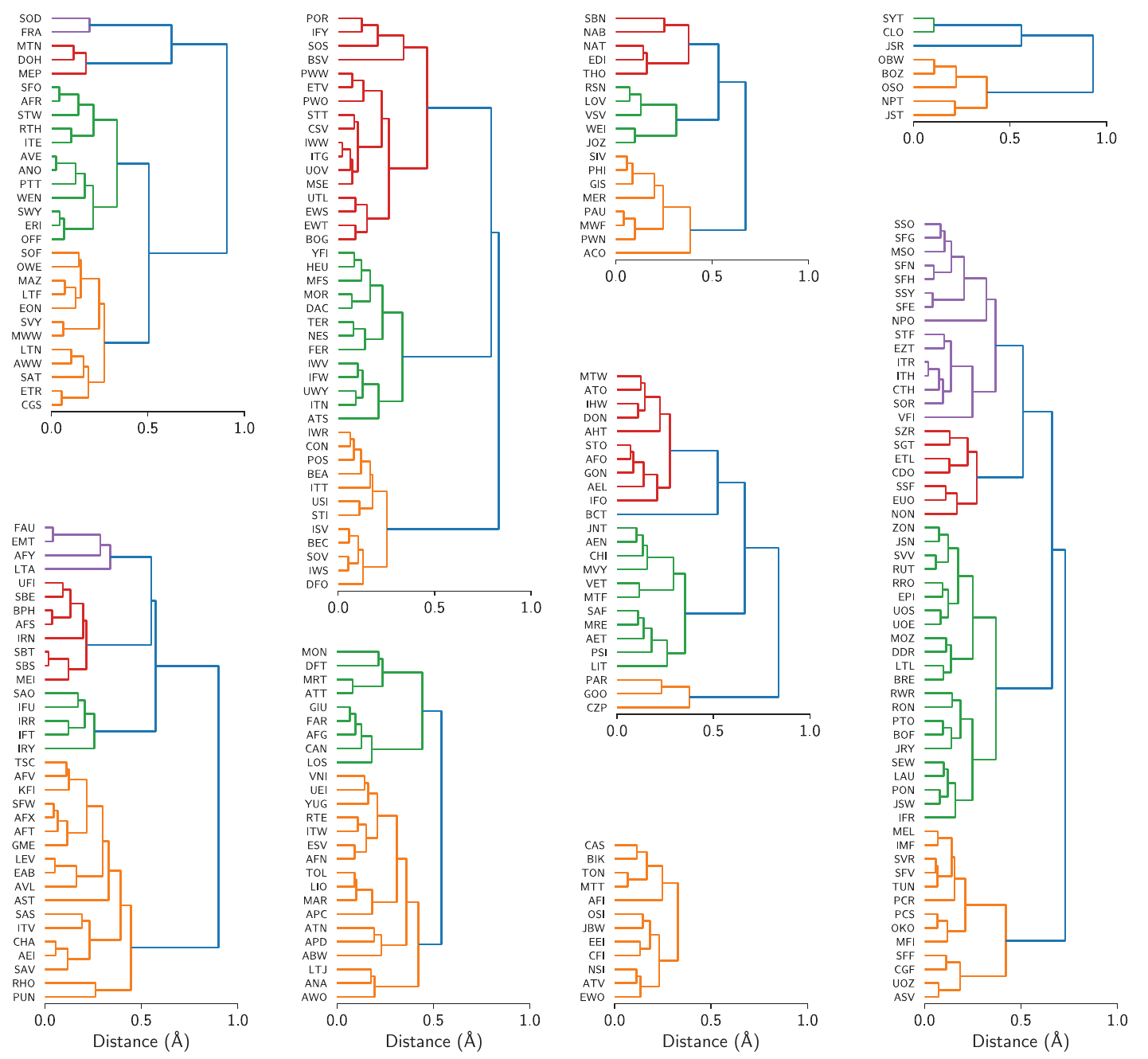}
    \caption{Subclusters of the full dendrogram in Fig. \ref{fig:si:03-dendrogram-full} with labels.}
    \label{fig:si:03-dendrogram-clusters}
\end{figure}

\begin{figure}[!h]
    \centering
    \includegraphics[width=\linewidth]{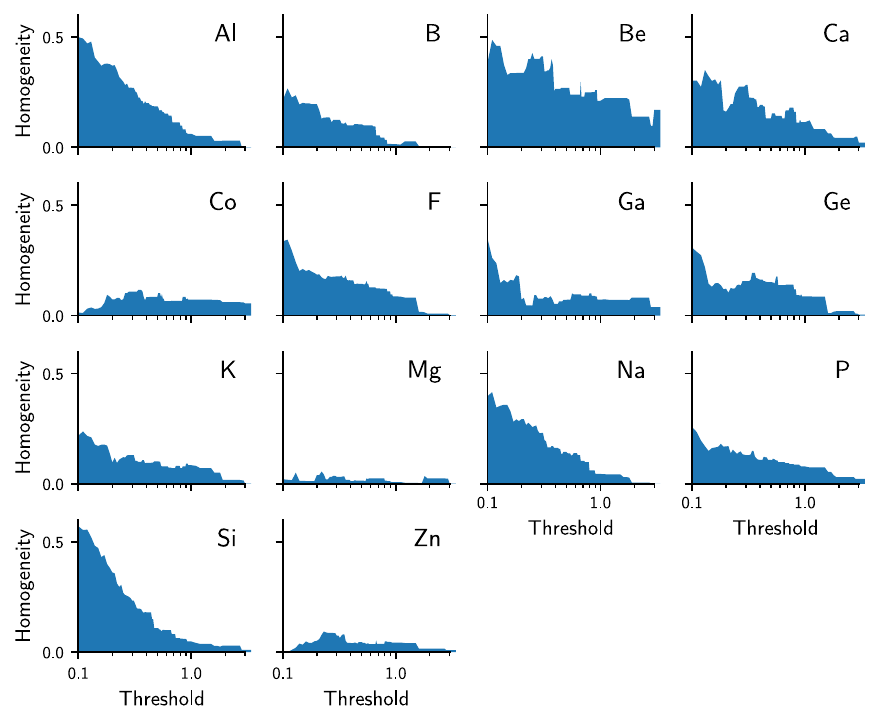}
    \caption{Clustering homogeneity for different synthesis conditions when flat clusters are created with the given threshold. A homogeneity of zero indicates a perfect mixing between positive and negative labels in the same clusters.}
    \label{fig:si:03-synthesis-homogeneity}
\end{figure}

\begin{figure}[!h]
    \centering
    \includegraphics[width=\linewidth]{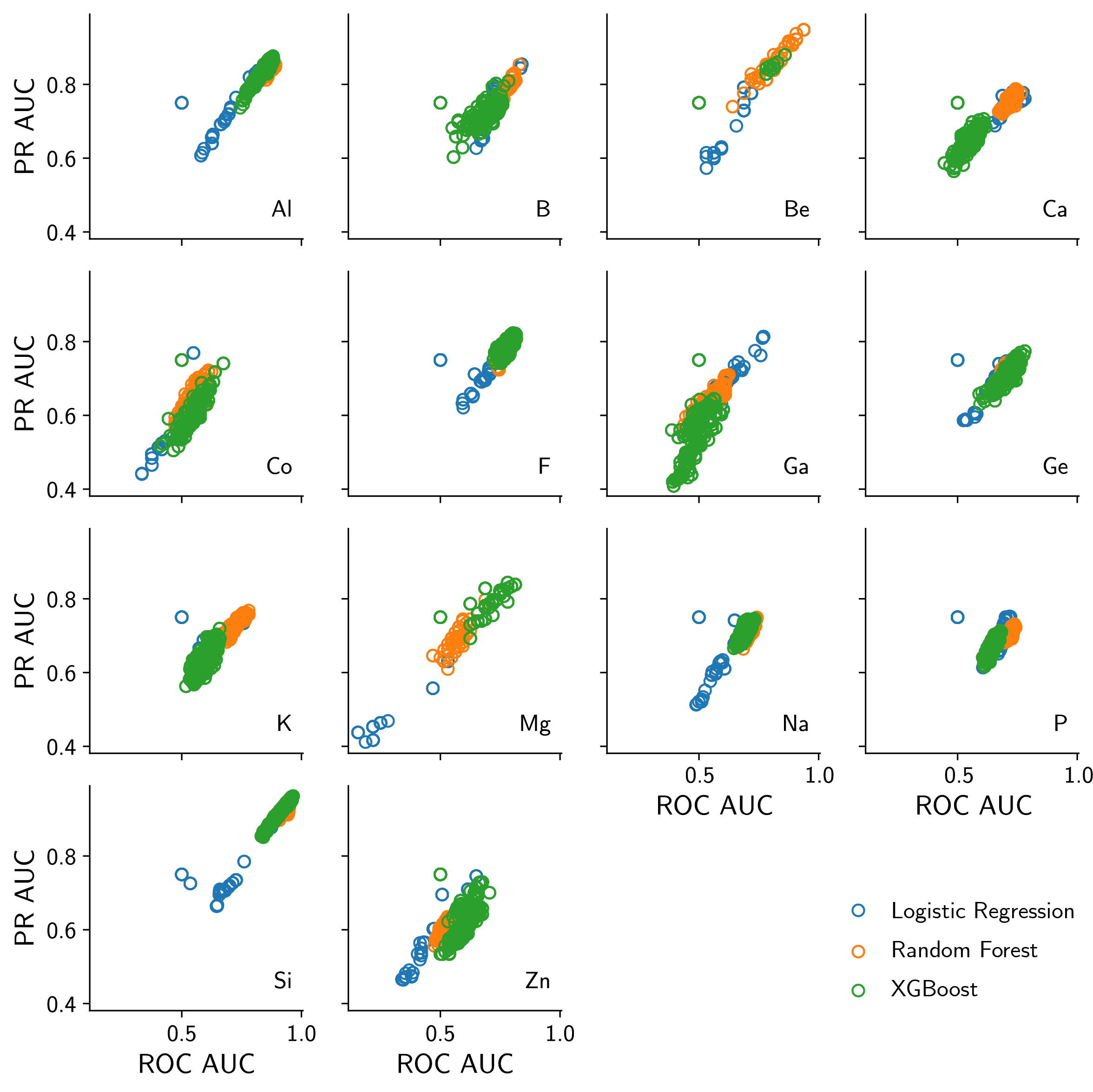}
    \caption{Precision-recall (PR) and Receiver Operating Characteristic (ROC) areas under the curve (AUC) for different hyperparameters of three different classifiers: logistic regression, random forest, and XGBoost. The PR AUC and ROC AUC are adopted as the main figures of merit for evaluating these classifiers.}
    \label{fig:si:04-classifiers-comparison}
\end{figure}

\begin{figure}[!h]
    \centering
    \includegraphics[width=\linewidth]{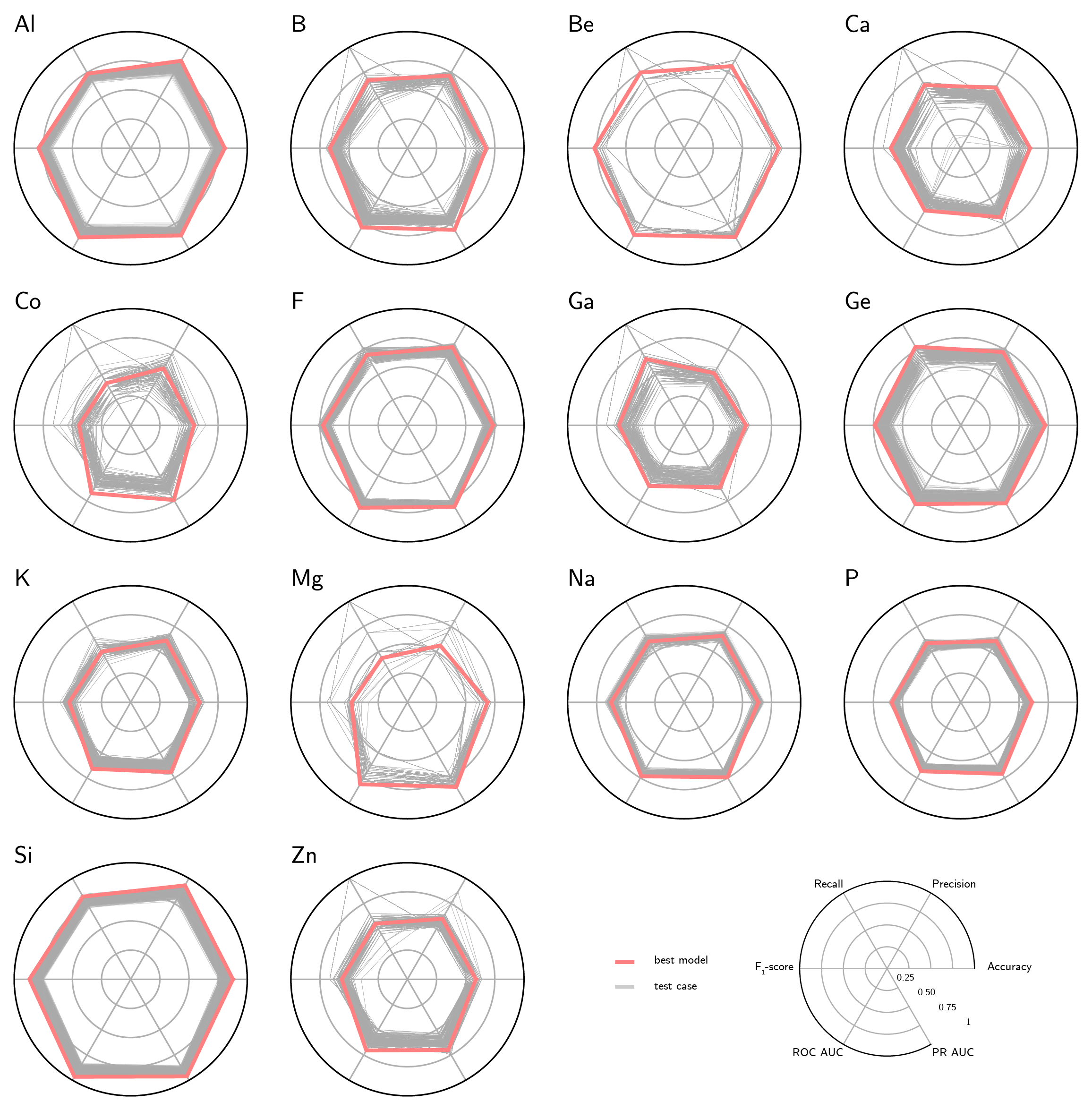}
    \caption{All figures of merit for different hyperparameters of the XGBoost classifiers. Each gray line represents the average metric of five runs at each set of hyperparameters. The figures of merit of interest, along with the values of each line in the circle, are shown in the lower right of the plot.}
    \label{fig:si:04-xgboost-metrics}
\end{figure}

\begin{figure}[!h]
    \centering
    \includegraphics[width=\linewidth]{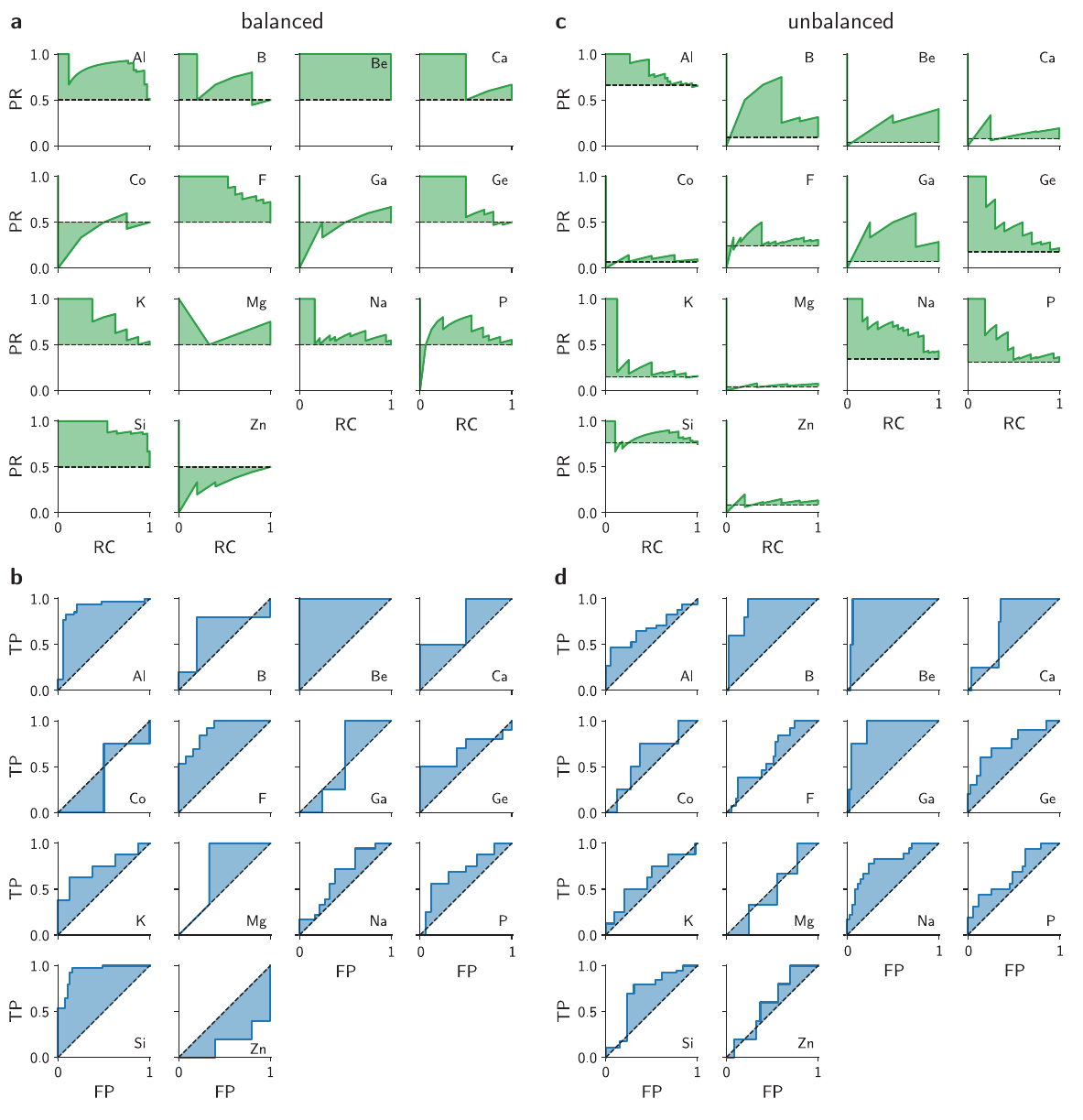}
    \caption{Precision (PR) and recall (RC) curves for the best \textbf{a}, balanced and \textbf{b}, unbalanced XGBoost classifier. The receiver operating characteristic curve, shown with the true positive (TP) and false positive (FP) ratios, are also depicted for a \textbf{c}, balanced and \textbf{d}, unbalanced XGBoost classifier. The best hyperparameters are selected according to the validation results. The curves in this figure are computed for a held-out test split.}
    \label{fig:si:04-xgboost-auc}
\end{figure}

\begin{figure}[!h]
    \centering
    \includegraphics[width=0.5\linewidth]{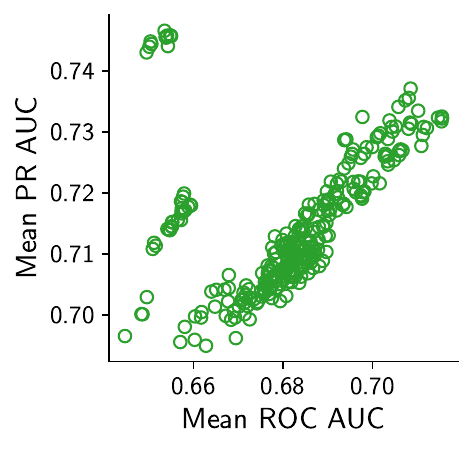}
    \caption{Mean area under the ROC and PR curves for the XGBoost classifiers, computed with respect to all synthesis predictions at once. The best models maximize both the PR and ROC curves.}
    \label{fig:si:04-xgboost-best}
\end{figure}

\begin{figure}[!h]
    \centering
    \includegraphics[width=\linewidth]{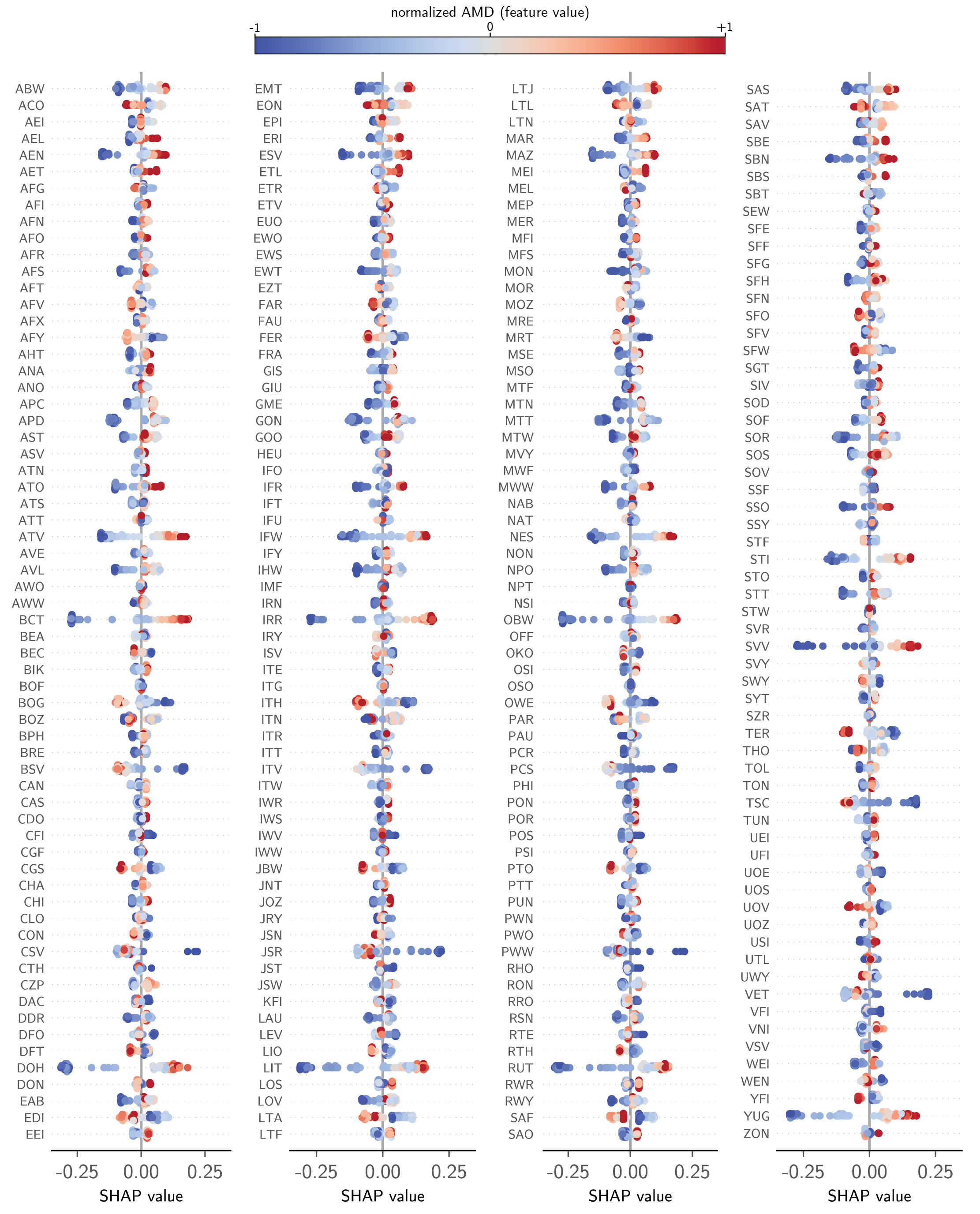}
    \caption{Visualization of the SHAP values for a single XGBoost classifier predicting whether Ge should be used in the synthesis of known zeolites. To plot the results in this graph, each feature (AMD distance to a given zeolite) was normalized in a per-feature basis. Larger SHAP values indicate higher impact in a positive classification.}
    \label{fig:si:05-classifier-shap}
\end{figure}

\begin{figure}[!h]
    \centering
    \includegraphics[width=\linewidth]{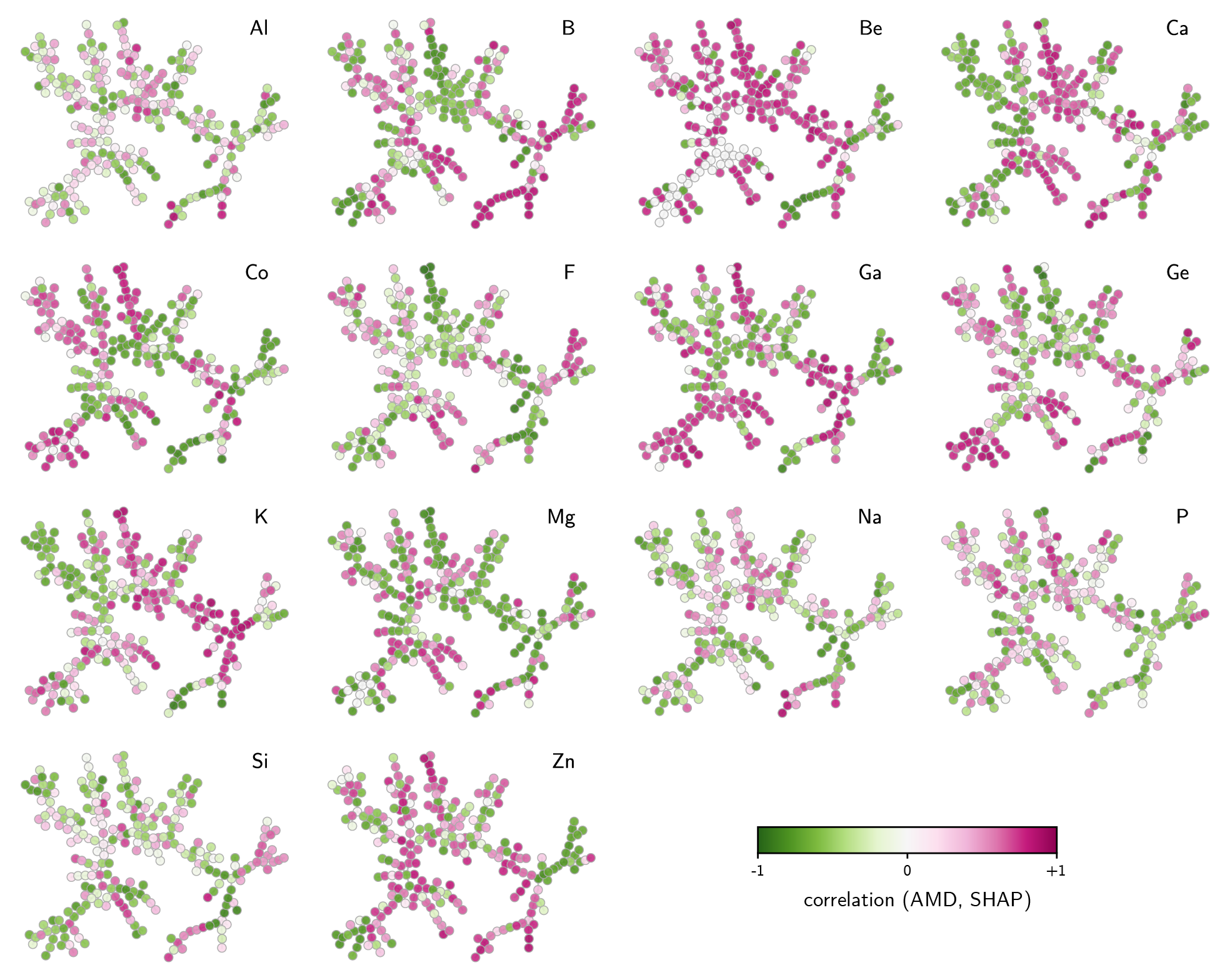}
    \caption{Zeolite tree map labeled with the Pearson correlation coefficient between AMD and SHAP values per synthesis conditions. A negative correlation (shown in green) indicates that smaller AMD distances (i.e., high similarity) lead to higher SHAP values (i.e., higher likelihood of a positive classification). The correlation coefficient is the average correlation of AMD and SHAP values for 100 XGBoost models for each synthesis condition.}
    \label{fig:si:05-classifier-explainability}
\end{figure}

\begin{figure}[!h]
    \centering
    \includegraphics[width=0.5\linewidth]{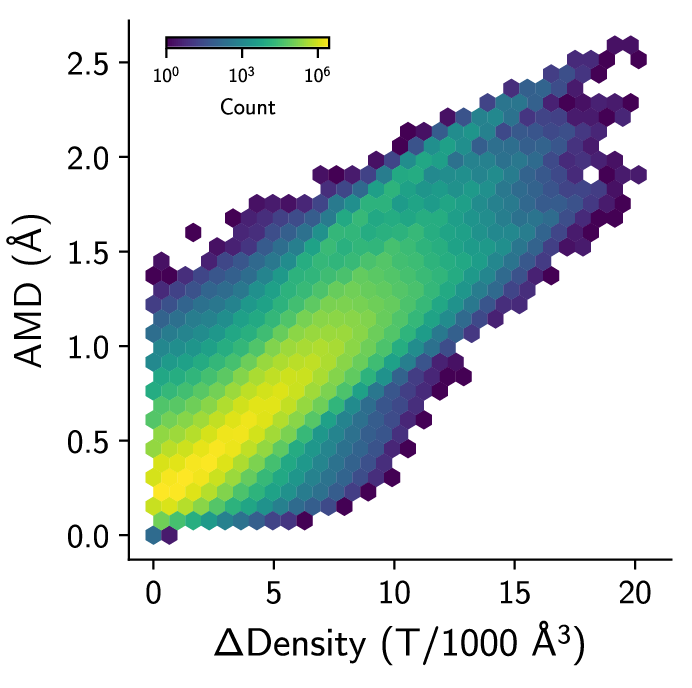}
    \caption{Relationship between AMD distances computed between known and hypothetical frameworks, and their density difference. Both known and hypothetical zeolites were optimized with the NNPscan method by Erlenbach et al.}
    \label{fig:si:06-hyp-analysis}
\end{figure}

\begin{figure}[!h]
    \centering
    \includegraphics[width=\linewidth]{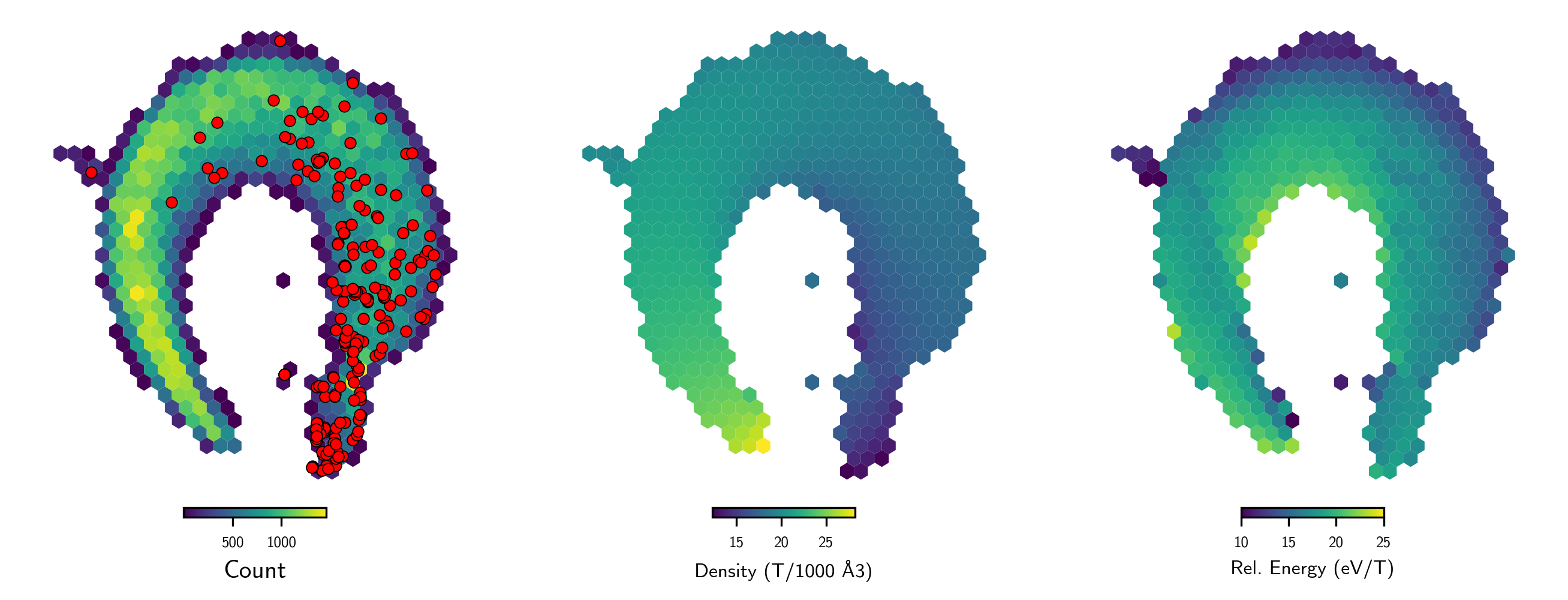}
    \caption{Low-dimensional projection of the hypothetical zeolite space using their distance towards known zeolites as features. The red dots indicate zeolites present in the IZA database. The low-dimensional plot was obtained using UMAP (see Methods), and recovers the density and energy of the frameworks simply by comparing them against known structures.}
    \label{fig:si:06-hyp-umap}
\end{figure}

\begin{figure}[!h]
    \centering
    \includegraphics[width=0.75\linewidth]{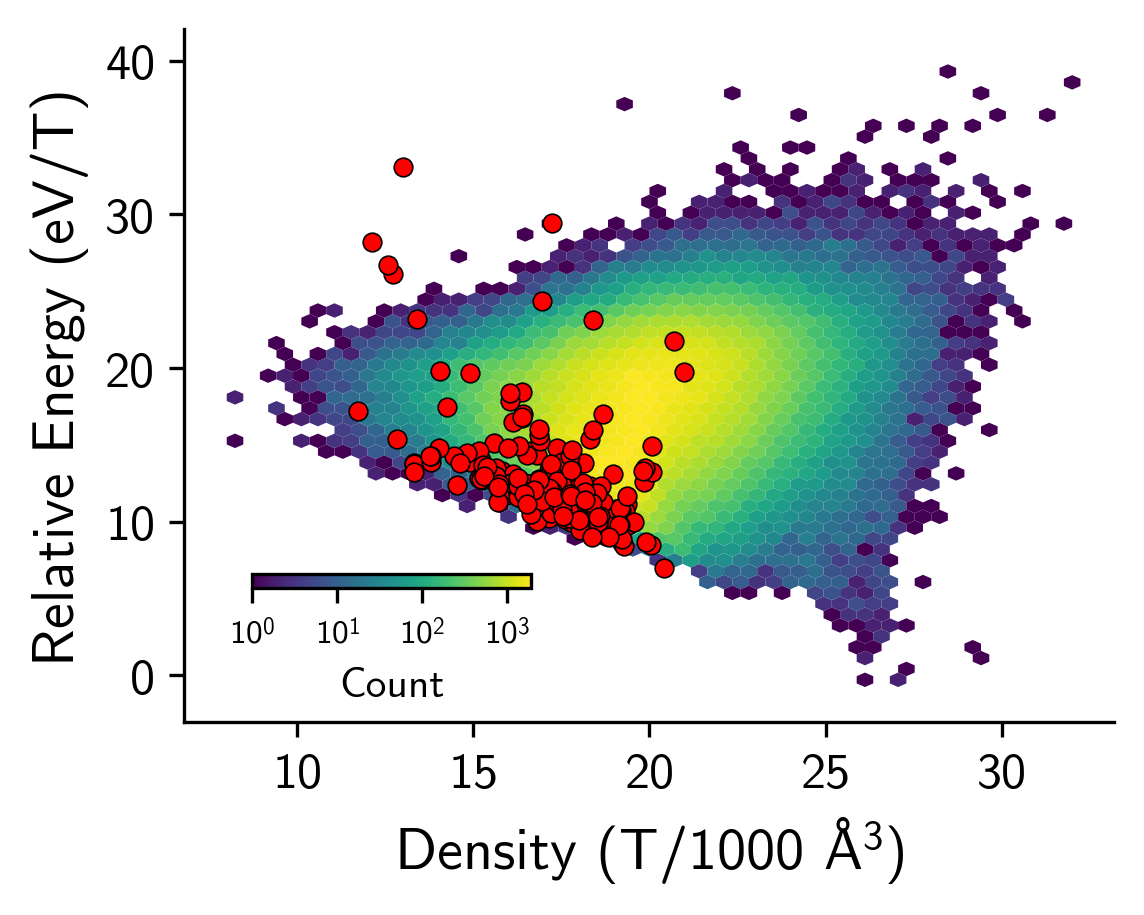}
    \caption{Energy-density plots for zeolites, reproducing the plot from Erlenbach et al. The analysis of the data using these two variables shows a high concentration of mid-energy, mid-density zeolites in the hypothetical structures database.}
    \label{fig:si:06-hyp-density}
\end{figure}

\begin{figure}[!h]
    \centering
    \includegraphics[width=\linewidth]{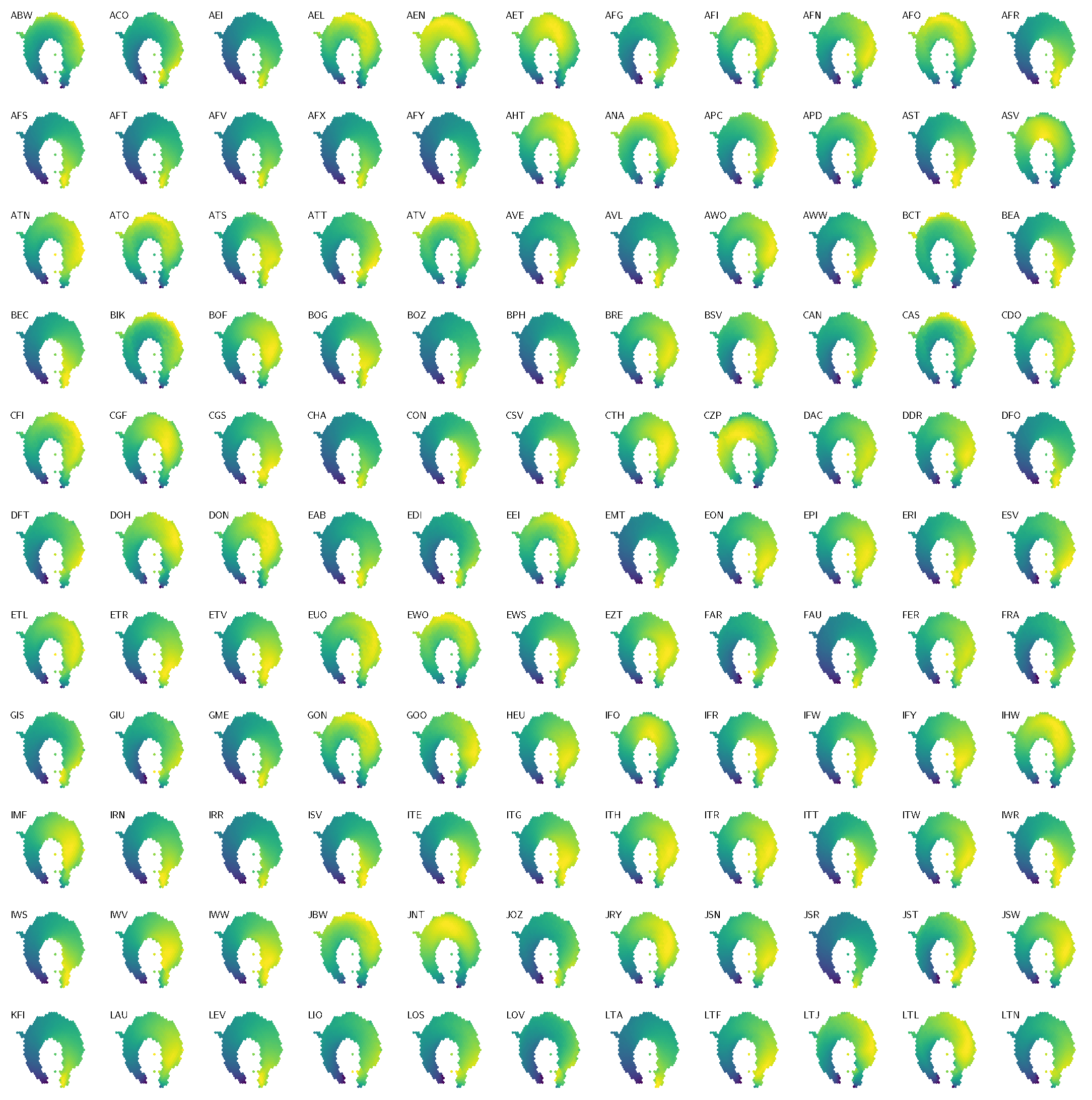}
    \caption{Average distance of hypothetical zeolites towards known zeolites, visualized using the UMAP plot from Fig. \ref{fig:si:06-hyp-umap}. Brighter colors indicate lower distances. (continues in Fig. \ref{fig:si:06-hyp-iza-02}).}
    \label{fig:si:06-hyp-iza-01}
\end{figure}

\begin{figure}[!h]
    \centering
    \includegraphics[width=\linewidth]{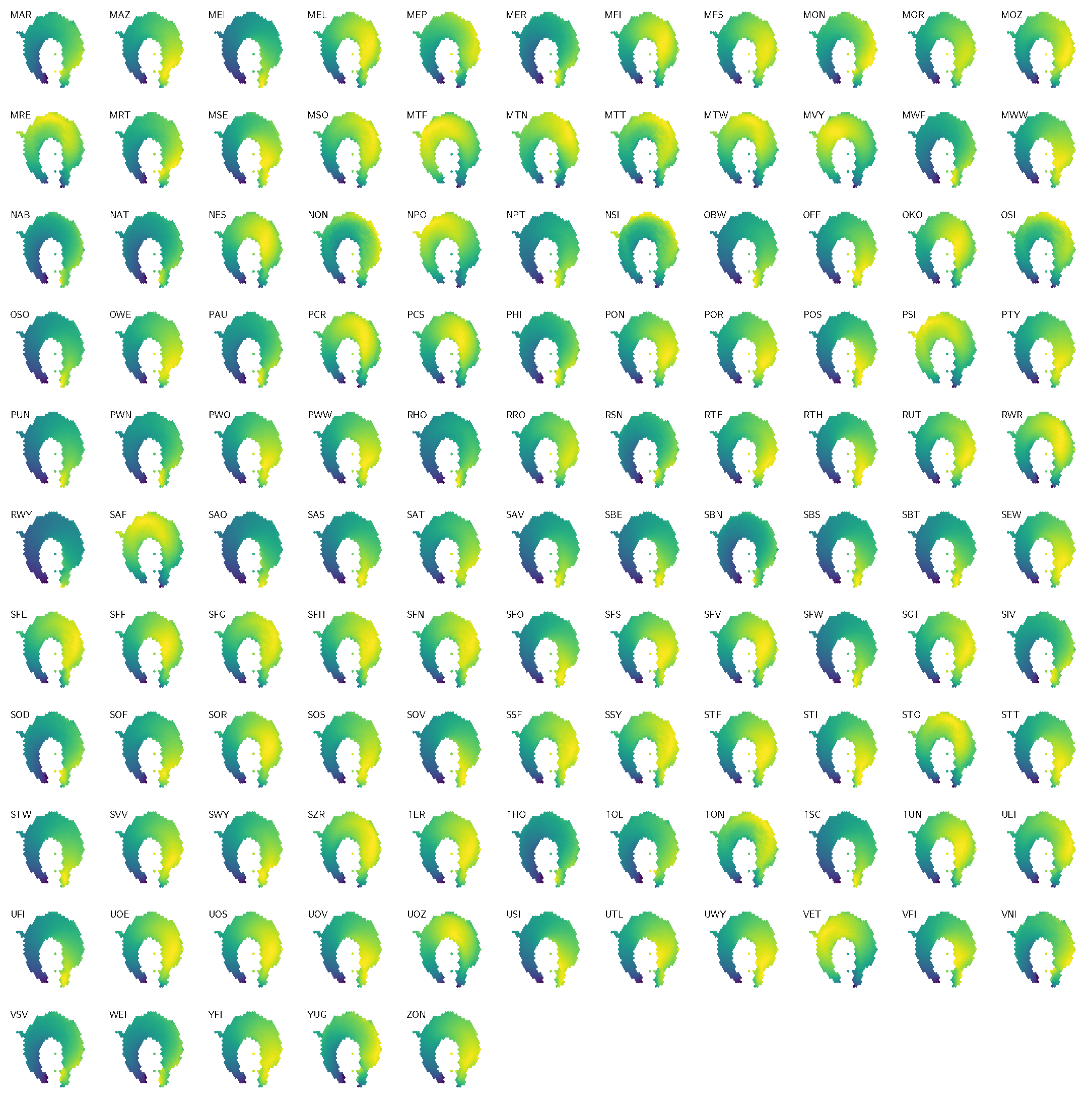}
    \caption{Average distance of hypothetical zeolites towards known zeolites, visualized using the UMAP plot from Fig. \ref{fig:si:06-hyp-umap}. Brighter colors indicate lower distances. (continued from Fig. \ref{fig:si:06-hyp-iza-01}).}
    \label{fig:si:06-hyp-iza-02}
\end{figure}

\begin{figure}[!h]
    \centering
    \includegraphics[width=\linewidth]{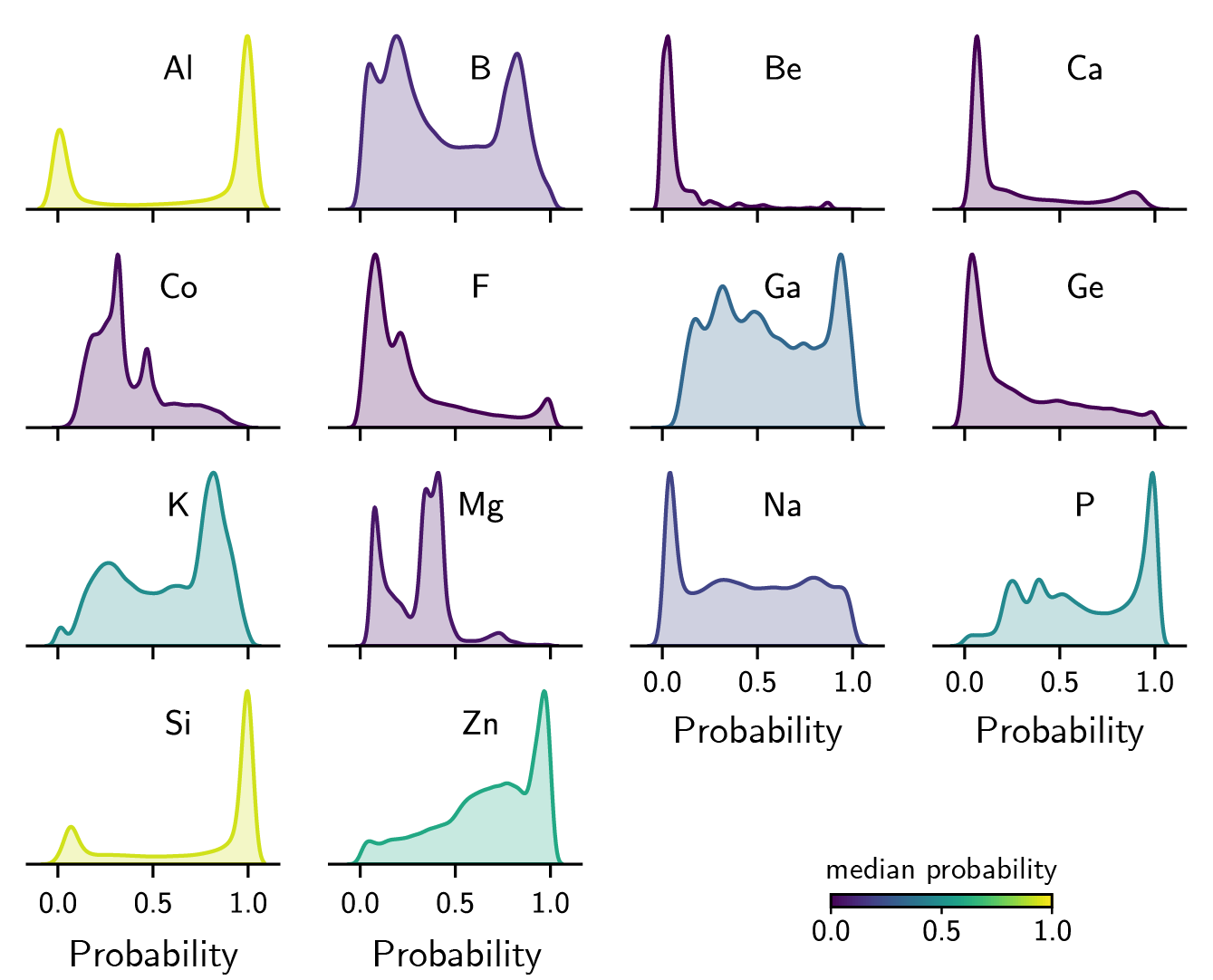}
    \caption{Distributions of synthesis probabilities, as predicted by 100 XGBoost models, for all hypothetical zeolites in the ``Deem dataset''. Brighter colors indicate higher medians of the distributions.}
    \label{fig:si:06-hyp-inorganics}
\end{figure}

\clearpage
\section{Supporting Tables}

\begin{table}[h]
\centering
\caption{Closest pairs of IZA zeolites according to AMDs. The pairs shown are sorted according to distances between their AMDs, and have maximum AMD distance of 0.05 \AA.}
\label{tab:si:iza_pairs}
\begin{tabular}{ccc}
\hline\hline
Zeolite 1 & Zeolite 2 & AMD distance (\AA) \\
\hline
ITH &      ITR &  0.016 \\
ITG &      UOV &  0.021 \\
SBS &      SBT &  0.022 \\
ITG &      IWW &  0.025 \\
MEL &      SFV &  0.032 \\
MWF &      PAU &  0.033 \\
IMF &      SFV &  0.035 \\
IMF &      TUN &  0.035 \\
AFG &      TOL &  0.037 \\
FAR &      MAR &  0.037 \\
ERI &      SWY &  0.039 \\
OFF &      SWY &  0.040 \\
AFT &      AFX &  0.040 \\
IWW &      UOV &  0.041 \\
AFS &      BPH &  0.042 \\
SFH &      SFN &  0.042 \\
AFX &      SFW &  0.044 \\
EMT &      FAU &  0.045 \\
LIO &      TOL &  0.045 \\
PTY &      PWO &  0.046 \\
IWS &      SOV &  0.047 \\
AWO &      UEI &  0.047 \\
ITG &      UWY &  0.050 \\
\hline\hline
\end{tabular}
\end{table}

\begin{table}[h]
\centering
\caption{Hyperparameters explored for the logistic regression models. For the $L_1$ loss, only the \texttt{saga} solver was used. The \texttt{l1\_ratio} parameter is used only in the case of the $L_1$ loss.}
\label{tab:si:logistic}
\begin{tabular}{ll}
\hline \hline
Parameter & Choices \\ \hline
\texttt{penalty} & [\texttt{l2}, \texttt{l1}, \texttt{none}] \\
\texttt{C} & [0.001, 0.01, 0.1, 1, 10, 100] \\
\texttt{solver} & [\texttt{lbfgs}, \texttt{liblinear}, \texttt{sag}, \texttt{saga}] \\
\texttt{l1\_ratio} & [0.25, 0.5, 0.75, 1.0] \\
\hline \hline
\end{tabular}
\end{table}

\begin{table}[h]
\centering
\caption{Hyperparameters explored for the random forest classifiers.}
\label{tab:si:rf}
\begin{tabular}{ll}
\hline \hline
Parameter & Choices \\ \hline
\texttt{n\_estimators} & [50, 100, 200] \\
\texttt{max\_depth} & [None, 10, 20] \\
\texttt{min\_samples\_split} & [2, 5, 10] \\
\texttt{min\_samples\_leaf} & [1, 2, 4] \\
\texttt{bootstrap} & [True, False] \\
\hline \hline
\end{tabular}
\end{table}

\begin{table}[h]
\centering
\caption{Hyperparameters explored for the XGBoost classifiers.}
\label{tab:si:xgb}
\begin{tabular}{ll}
\hline \hline
Parameter & Choices \\ \hline
\texttt{n\_estimators} & [50, 100, 200] \\
\texttt{learning\_rate} & [0.01, 0.1, 0.2] \\
\texttt{max\_depth} & [3, 4, 5, 6] \\
\texttt{min\_child\_weight} & [1, 2, 3] \\
\texttt{subsample} & [0.5, 0.75, 1] \\
\texttt{colsample\_bytree} & [0.5, 0.75, 1] \\
\hline \hline
\end{tabular}
\end{table}

\end{document}